\documentclass[twocolumn]{aastex63}
\usepackage[version=4]{mhchem}
\usepackage{booktabs} 
\usepackage{graphics, multirow}
\newcommand{\cmmnt}[1]{}
\usepackage{hyperref} 
\usepackage{graphics, multirow}
\usepackage{enumitem}
\usepackage{refcount}
\usepackage[ugly]{nicefrac}
\usepackage{float} 

\graphicspath{ {./figures/} }

\shorttitle{ Molecular Absorption Cross-Section Database for Exoplanets/Brown-dwarfs}
\shortauthors{Gharib-Nezhad et al.}

\usepackage{subfigure}
\begin{document}

\title{{\tt EXOPLINES}: Molecular Absorption Cross-Section Database for Brown Dwarf and Giant Exoplanet Atmospheres}

\author[0000-0002-4088-7262]{Ehsan Gharib-Nezhad}

\correspondingauthor{Ehsan Gharib-Nezhad} \thanks{NASA Postdoctoral Fellow}
\email{e.gharibnezhad@asu.edu}
\affiliation{NASA Ames Research Center, Moffett Field, CA 94035, USA}
\affil{Bay Area Environmental Research Institute, NASA Ames Research Center, Moffett Field, CA 94035, USA}
\affil{School of Molecular Sciences, Arizona State University, Tempe, AZ 85281, USA}

\author{Aishwarya R. Iyer}
\affiliation{School of Earth and Space Exploration Arizona State University, Tempe AZ 85281, USA}

\author{Michael R. Line}
\affiliation{School of Earth and Space Exploration Arizona State University, Tempe AZ 85281, USA}

\author{Richard S. Freedman}
\affiliation{SETI Institute, Mountain View, CA 94035, USA}
\affiliation{NASA Ames Research Center, Moffett Field, CA 94035, USA}

\author{Mark S. Marley}
\affiliation{NASA Ames Research Center, Moffett Field, CA 94035, USA}

\author{Natasha E. Batalha}
\affiliation{NASA Ames Research Center, Moffett Field, CA 94035, USA}

\begin{abstract}
Stellar, substellar, and planetary atmosphere models are all highly sensitive to the input opacities. Generational differences between various state-of-the-art stellar/planetary models are primarily because of incomplete and outdated atomic/molecular line-lists. Here we present a database of pre-computed absorption cross-sections for all isotopologues of key atmospheric molecules relevant to late-type stellar, brown dwarf, and planetary atmospheres: \ce{MgH}, \ce{AlH}, \ce{CaH}, \ce{TiH}, \ce{CrH}, \ce{FeH}, \ce{SiO}, \ce{TiO}, \ce{VO}, and \ce{H2O}. The pressure and temperature ranges of the computed opacities are between 10$^{-6}$--3000~bar and 75--4000~K, and their spectral ranges are 0.25--330~$\mu$m for many cases where possible. For cases with no pressure-broadening data, we use collision theory to bridge the gap. We also probe the effect of absorption cross-sections calculated from different line lists in the context of Ultra-Hot Jupiter and M-dwarf atmospheres. Using 1-D self-consistent radiative-convective thermochemical equilibrium models, we report significant variations in the theoretical spectra and thermal profiles of substellar atmospheres. With a 2000 K representative Ultra-Hot Jupiter, we report variations of up to 320 and 80 ppm in transmission and thermal emission spectra, respectively. For a 3000 K M-dwarf, we find differences of up to 125$\%$ in the spectra. We find that the most significant differences arise due to the choice of TiO line-lists, primarily below 1$\mu$m. In sum, we present (1) a database of pre-computed molecular absorption cross-sections, and (2) quantify biases that arise when characterizing substellar/exoplanet atmospheres due to line list differences, therefore highlighting the importance of correct and complete opacities for eventual applications to high precision spectroscopy and photometry.
\end{abstract}



\section{Introduction}\label{sec:intro}
Over the past two decades, atmospheric characterization of exoplanets and brown dwarfs has been dramatically increasing---particularly, observations with $HST$, $Spitzer$, various ground-based telescopes both with low to moderate resolution spectroscopy and high-resolution cross-correlation approaches \citep[e.g.,][]{Madhusudhan2019}. Additionally, with the Transiting Exoplanet Survey Satellite ($TESS$), in conjunction with the launch of the James Webb Space Telescope ($JWST$), we can expect hundreds of planet detections \citep{Sullivan2015, Louie2018, Kempton2018TESS} followed by detailed spectroscopic characterization of their atmospheres \citep{Greene2016JWST, Bean2018JWST}. The compositional and temperature diversity of these worlds provide a unique opportunity to gain insight into planetary atmospheric chemical processes and formation pathways \citep{Oberg2011, Fortney2013, Mordasini2016,Horst2018Haze}.

Fully capitalizing on the information encoded in a spectrum, therefore, requires a robust model interpretation. Models used to interpret thermal emission spectra (either from an isolated object like a brown dwarf, M-dwarf, or directly imaged planet, or a secondary eclipse spectrum for a transiting planet) and transmission spectra (for transiting planets) require some combination of chemistry and radiative transfer. Self-consistent models (either 1D or 3D) rely upon a detailed coupling between chemical processes (e.g., thermochemical equilibrium or chemical kinetics/photochemistry), radiative energy transport, convective/dynamical energy transport, and cloud/haze processes to determine the atmospheric state and subsequent observables \citep[e.g.,][and references therein]{Marley2015,Zhang2020}. Retrieval models, on the other hand, are purely data-driven and attempt to leverage large numbers of free parameters (e.g., molecular abundances for selected species, temperature profile parameterization) to adequately fit the observables to determine Bayesian constraints on composition, clouds, and thermal structures \citep{Line2014}. The adequacy of these methods, whether for predictions or parameter estimation, relies upon accurate and complete molecular and atomic opacities.

In this work, we focus on high-resolution (adaptive R, e.g., 2$\times$10$^4$--2$\times$10$^6$ at 1~$\micron$) opacities with implications to Ultra-Hot Jupiter atmospheres. Our prime candidates being \ce{TiO}, \ce{VO}, \ce{MgH}, \ce{AlH}, \ce{CaH}, \ce{TiH}, \ce{CrH}, \ce{FeH}, and \ce{SiO}--the key molecules{\footnote{For further study on the composition of these ultra-hot Jupiters and M-dwarfs, studies by \citet{Lodders2006LMS} and \citet{Visscher2010} are suggested.}} in such atmospheres. In addition, we generate \ce{H2O} high-resolution opacities from 500--4000~K, which is required for modeling everything from super-Earth to M-dwarf atmospheres. Starting with \S\ref{sec:astroImportance}, we first provide an astrophysical context for the importance of metal hydrides and oxides in substellar atmospheres. \S\ref{sec:XS-background} provides discussion on the current state of opacity or absorption cross-section data, highlighting line broadening effects that are missing and how we tackle them. In \S \ref{sec:Results}, we bridge the gap by providing pre-generated absorption cross-sections for the metal hydrides and oxides in question, alongside comparisons with previous efforts. In \S\ref{sec:Impact-on-RTmodeling}, we quantify the effect of opacity differences on transmission and emission spectra of a representative\footnote{``representative'' in this context means a generic atmospheric case} Ultra Hot Jupiter planet, followed by M-dwarf stellar spectra in \S\ref{sec:Impact-on-RTmodeling-Mdwarf}, and potential biases and challenges with opacities that arise in atmospheric characterization efforts, described in \S\ref{sec:opacity-challenge}. Finally, we present a brief summary of our conclusions in \S\ref{sec:conclusion}. The main outcome of this work is {\tt EXOPLINES} absorption cross-section database (i.e., EXOplanet Pressure-broadened LINES).

\section{Astrophysical Importance}\label{sec:astroImportance}

Temperature and pressure play a key role in controlling molecular and atomic abundances that ultimately shape a spectrum \citep[e.g.,][]{burrows1999chemical}. At cooler temperatures ($<$200~K), solar system giants are dominated by absorption because of methane and scattering by clouds, and photochemically produced hydrocarbons. At warmer temperatures (300--1000~K), T-dwarfs, and many (sub)Neptune transiting planet atmospheres are sculpted (or presumed to be sculpted) by water, methane, ammonia, and alkali metals. Higher temperature results in the onset of CO absorption, followed by several oxides (TiO, VO) and hydrides (FeH, CrH, MgH, CaH), especially above 1500~K. The increasing presence of these species as a function of temperature is what ultimately (along with dust/condensate formation) governs the M-L-T-Y dwarf classification sequence and their sub-types \citep{Kirkpatrick2005}. Furthermore, water dissociation, the onset of H-bound/free-free, and contribution of refractory atoms and their ions (e.g., Fe, Ti, Mg, Ca, Si) dominate atmospheres with temperatures above $\sim$2500~K, encountered in M-dwarfs and ultra-hot Jupiters \citep{Lodders1999chem}. Here, we highlight the importance of these metal oxides and hydrides and their behavior in high-temperature regimes---particularly in substellar atmospheres of M/L-type Dwarfs and highly irradiated exoplanets.

\subsection{Metal Oxides}

M-dwarf stars {(2000 K $<$ T$_\mathrm{eff}$ $<$ 4000 K)}, active photospheric spots on sun-like stars, and even highly irradiated exoplanets with similar effective temperatures have atmospheres that are all dominated by TiO \citep{Kirkpatrick1991, Webber1971}. Covering the optical spectral region between 630--900~nm, TiO plays a vital role in the proper characterization of M-dwarfs and Ultra-Hot Jupiters. The temperature dependence of TiO governs the energy level populations that affect the relative band-structure intensities, an essential tool in classifying early-type M-dwarfs \citep{Sharpless1956}. For instance, in spectra of M0--M3 dwarfs, the 0--0 band\footnote{In spectroscopy, the $\nu^\prime-\nu^{\prime\prime}$ band refers to all rotation-vibrational transitions between the upper vibrational level, $\nu^\prime$, and the lower vibration level $\nu^{\prime\prime}$ \citep{Bernath_book}.} (705.4~nm) is used as the primary classification marker. The strongest TiO bands near 843~nm,  819.4~nm to 885.9~nm, and 660 to 710~nm are important indicators for M4, M6, and late-type M8 dwarfs respectively \citep{Sharpless1956, Lockwood1972, Ruiz1997,Delfosse1997}.

Similar to TiO, VO is a refractory molecule in M-dwarfs and early L-dwarf atmospheres ($\sim$L2). With strong spectral signatures between 0.6--1.5~$\micron$, metal oxides such as VO are sensitive temperature indicators in substellar atmospheres \citep{Cushing2005}. The strength of the VO absorption features at 1.17 and 1.22~$\micron$, combined with TiO and CaH bands, are valuable spectroscopic youth indicators for brown dwarfs \citep{Kirkpatrick2006, Peterson2008}.

In addition to their importance for substellar spectral and age classification, high-temperature oxides are critical sources of opacity in the optical wavelength region, that drive heating in the upper atmospheres of hot Jupiters ($>$1800~K and $<$50~mbar)---essential for proper characterization of the 1D vertical structure of the atmosphere \citep{Hubeny2003Irradiated, Fortney2008UHJ-TiO-VO}. The presence of TiO and VO drive up stellar irradiation mean opacity relative to the planetary thermal emission mean opacity, resulting in a ``thermal inversion" (increasing temperature with decreasing pressure) in the planetary atmosphere. However, we note that bonafide detection of metal oxides (TiO or VO) has remained elusive and somewhat controversial, partially due to the inadequacy of the opacity data for these species \citep{Merritt2020NoDetection, Nugroho2017TiO-WASP33b}.

In substellar atmospheres, especially at temperatures cooler than 2300 K, both TiO and VO gradually disappear due to condensate formation (TiO into to \ce{CaTiO3} and VO turns into solid VO, \cite{Lodders1999chem}). Therefore, their spectral features are not prominent at such atmospheres at $\sim$0.7--1.2 ~$\micron$, allowing for metal hydride FeH, CrH, and CaH features to present themselves.

\subsection{Metal Hydrides}
    
Several FeH features within the 0.7--1.3~$\micron$ range has been detected in the late-M and L dwarfs \citep{Kirkpatrick2000}, including the 0--0 band (known as the Wing-Ford band) at 0.985--1.02~$\micron$ \citep{Cushing2003FeH}, the 1--0 band at 869.2~nm in L-dwarfs \citep{Kirkpatrick1999_SpectralTypeL}, the 2--0 band at 778.6~nm, the 1--0 band at 869.2~nm, the 2--1 band at 902.0~nm \citep{Tinney1998FeH-VO} and the 1--2 band at 1238.9~nm bands are present in very low-mass stars \citep{Reid2001dwarfsFeH}. Additionally, \citet{Schiavon1997FeH} showed that the Wing-Ford band is sensitive to surface gravity and metallicity. Similar to FeH, CaH, CrH, and AlH all show a strong dependence on surface gravity (see Chapter 4 in \citet{Reid2005book}).

The CrH 0--0 (861.1~nm) and 0--1 (996.9~nm) bands along with the FeH bands (and their ratios) serve as diagnostic indicators \citep{Kirkpatrick1999_SpectralTypeL,Kirkpatrick2000} for classifying the L-dwarf sub-types (see Table 6 in \citet{Kirkpatrick1999_SpectralTypeL}).

In the context of Ultra-Hot Jupiters, species like MgH, AlO, NaH, and CaO could also drive thermal inversions depending on elemental abundance ratios \citep{Gandhi2019ThermalInversions}. Additionally, hydrides such as MgH, CrH, CaH, and FeH although expected to be present in hot Jupiter atmospheres; the lack of reliable detections have made it difficult to assess their true role in governing the 1-D thermal structure.

\section{Methodology: Molecular Absorption Cross-Sections} \label{sec:XS-background}
The absorption cross section\footnote{Both opacities and absorption cross-sections represent the capability of species (i.e., atom, molecule, or radical) to absorb or emit light. However, their definitions are different. Opacity $\kappa$ is the ``effective area’’ of one gram of an absorber (unit: cm$^{2}$/gram) to absorb/emit light while absorption cross-section $\sigma$ is the ``effective area’’ of one absorber (unit: cm$^{2}$/molecule). These quantities are interchangeable with $\kappa \rho$=$\sigma n$ equation, where $\rho$ is mass density ($gram/cm^3$) and $n$ is number density (molecule/cm$^3$).} (hereafter ACS) of an absorber is governed by the wavelength-dependent intensity, position, and width of a spectral line. The line shape is typically modeled with a Voigt profile, which is a convolution of Doppler (thermal) and pressure (Lorentz or collisional) broadening. Uncertainties in the line positions, strengths, and widths are ultimately what drive the uncertainties or inadequacies in various pre-computed ACSs. Below we present the formulation of line profile in \S\ref{sec:Line Width}, and then we extend our discussion on pressure-broadening and their availability in \S\ref{sec:Estimating-the-Lorentz-Coefficient}, followed by the treatment for their $J$-dependency and wing cut-off in \S\ref{sec:fitting-P-broadening}-\ref{sec:wing Cut-off}.
Finally, \S\ref{sec:linelist} describes the line-list data structure, and computational details are provided in \S\ref{sec:computational-details-ACS}.

\subsection{Line-width}\label{sec:Line Width}
{The conventional form of the spectral line shape is the Voigt profile, which is a convolution of Doppler and Lorentz profiles. Doppler broadening is governed by the temperature, molecular mass, and also the transition energy, and its half-width at half-maximum width (HWHM), ${\it \Gamma}_{\mathrm{D}}$, may be given by }\citep{Buldyreva2011book}:

\begin{equation} \label{eq:doppler}
	\Gamma_\mathrm{D} = 3.581 \times 10^{-7} \sqrt{\frac{T}{M}} \ \nu
\end{equation}

where $M$ is absorbers molecular mass in amu (atomic mass unit), $T$ is the temperature (in K), and $\nu$ is the line position in wavenumber (cm$^{-1}$). 

{ Lorentz line profile is controlled by the magnitude of the interaction between the absorber and broadening gas species, and the Lorentz HWHM, ${\it \Gamma}_{\mathrm{L}}$, depends on the temperature and the broadening species and can be described by }

\begin{equation}\label{eq:Lorentz-HWHM}
\centering
\Gamma _{\mathrm L} =  \sum _b \Big (\frac{T}{296 \mathrm{K}}\Big )^{-n_{T,b}} \;\gamma _{{\mathrm L,b}}\;p_{\mathrm{b}}
\end{equation}

where $\gamma_{\mathrm {L,b}}$ is the Lorentz coefficient of an absorber (HWHM, unit: cm$^{-1}$/bar), $p_{\mathrm{b}}$ is the broadener partial pressure (unit: bar), and $n_{\mathrm T,b}$ is the temperature-dependence coefficient, which is dimensionless. Both $\gamma_{\mathrm {L,b}}$ and  $n_{\mathrm T,b}$ are controlled by broadener--absorber interaction and the quantum properties of absorbing molecules. 

Quantifying $\gamma_{\mathrm L,b}$ and  $n_{\mathrm T,b}$ coefficients for the metal hydrides/oxides in question (e.g., TiO, VO, MgH, etc.) are challenging aspects of ACS calculations because there is not any laboratory or computational measurements have been done to determine their $\gamma_{\mathrm L,b}$ and  $n_{\mathrm T,b}$ coefficients\footnote{More precisely, we could find only one spectroscopic measurement of AlH in \ce{H2} gas by \citet{Watson1936}. Their measured $\gamma_\mathrm{L}$ is 0.24 cm$^{-1}$/atm, which looks much larger than the available \ce{H2}-broadening data for other well-known absorbers. Given the limitation in pressure gauges, spectrometers, and thermometers in their measurements in the 1930s, we decided not to use their number.}.

\subsection{Estimating the Lorentz Coefficient}\label{sec:Estimating-the-Lorentz-Coefficient}

Theoretically determining the Lorentz coefficient is quite challenging because it depends on molecular quantum properties such as total rotational angular momentum\footnote{In fact, each transition has its corresponding rotational and vibrational quantum numbers and molecular symmetries, which are controlling the Lorentz coefficient and hence the spectral line-width.}, molecular symmetries, and environmental conditions like temperature and the background gas broadener composition. For TiO, VO, TiH, CrH, the Lorentz coefficients have been estimated up to low $J$ quantum numbers (given in Table \ref{tab:Lorentz-coefficient-known-data}) for \ce{H2}/He atmospheres. These estimations are based on the similarity between the molecular symmetry or dipole moments that these absorbers have with the available Lorentz coefficients \citep[e.g.,][]{Freedman2014,Chubb2020}\footnote{For instance, \citet{Chubb2020} used HCN broadening coefficients for computing the ACS data of CaH, CrH, SiO, TiO, and VO absorbers, as well as HF broadening coefficients data for FeH and TiH.}).  In reality, we do not have enough or even a single measurement for the metal hydries/oxides in this study (except \ce{H2O})  at any temperature (see \S\ref{sec:opacity-challenge} for the challenges we are dealing with these opacities).

Due to the lack of $\gamma_{\mathrm L,b}$ and  $n_{\mathrm T,b}$ coefficients, we used the classical collision theory by \citet{Anderson1949} in order to estimate the Lorentz coefficient using the following equation (see \S\ref{sec:A1} for further discussion):

\begin{equation}\label{eq:LorentzCoeffCrossSection_cm-1/bar}
    \gamma_{\mathrm{L,b}} =  0.0567 \  (T \mu)^{-1/2} \ \sigma_{ \mathrm{col, b}}
\end{equation}

where $\mu$ is the reduced mass of the absorber-broadener pair in amu, $\sigma_{\rm{col, b}}$ is the collision cross-section in $\AA^2$ of the broadener b, and $\gamma_{\mathrm{L,b}}$ is in cm$^{-1}$/atm. In this equation, the collision (or scattering) cross section, $\sigma_{\rm{col, b}}$ is a fundamental factor, that connects collisions in the microscopic world into the spectral linewidth. For further details, see studies by \citet{Odashima1989} and \citet{Gierszal1998} that have used this approach to calculate the collision cross section from the laboratory-measured Lorentz line width. In addition, \citet{Cappelletti2005collision} have shown a good agreement between the experimental and computed pressure-broadening coefficients of \ce{C2H2}--Ar system.

{ Even though Eq.~\ref{eq:LorentzCoeffCrossSection_cm-1/bar} is derived from the collision theory, it has some limitations and errors such as ideal gas assumption and also possible inaccuracy in the computed collision cross-section data from ab-initio methods. Hence, this equation and the calculated pressure-broadening coefficients will not fulfill the lack of laboratory measurements or theoretical calculations. For example, the modified semi-classical Robert Bonamy theory \citep{Bonamy1979, Ma2007} have been successfully implemented to calculate the half-width and line shift of \ce{H2O} and CO absorbers in different broadening gases \citep[e.g.,][]{gamache2018temperature, VISPOEL2020H2O-N2}, and so similar studies are required for these molecules in question.

Note that the assumption of 0.5 for the temperature-dependence coefficient, $n_{\mathrm{T,b}}$ at Eq.~\ref{eq:LorentzCoeffCrossSection_cm-1/bar} is based on the kinetic theory, while laboratory measurements of other species suggest $n_{\mathrm{T,b}}$ can vary from 0.2--1.2. In turn, $n_{\mathrm{T,b}}$ can strongly depend on the broadener and $J$ quantum number (e.g., \ce{CH4} in different broadening gas that have shown in table 5 by \citet{GharibNezhad2019CH4}), or $n_{\mathrm{T,b}}$ may be roughly constant with $J$ (e.g., CO-self/air broadening by \citet{Devi2012}). \citet{Wagner2005_H2O-nT} have measured the effect of \ce{H2O}-air mixture and found that $n_{\mathrm{T,b}}$ can also be negative number. Table \ref{tab:collisionXS} represents the collected collision cross sections, $\sigma_{col}$ for CaH and SiO absorbers for \ce{H2} or He broadeners where available for $J_{\mathrm{lower}}$=0. In addition, the AlH $\gamma_{\mathrm{L,b}}$ coefficient is assumed to be equal to the CO broadening coefficient because of their similar dipole moments (we adopt this approach from \citet{Freedman2014} and \citet{Chubb2020} approaches}.)

 \subsection{Estimating the Behavior of Pressure-Broadening Coefficients and Their Dependence on $J$ Quantum Numbers}\label{sec:fitting-P-broadening}
 
\begin{figure*}[!b]
\centering
\includegraphics[width=\textwidth]{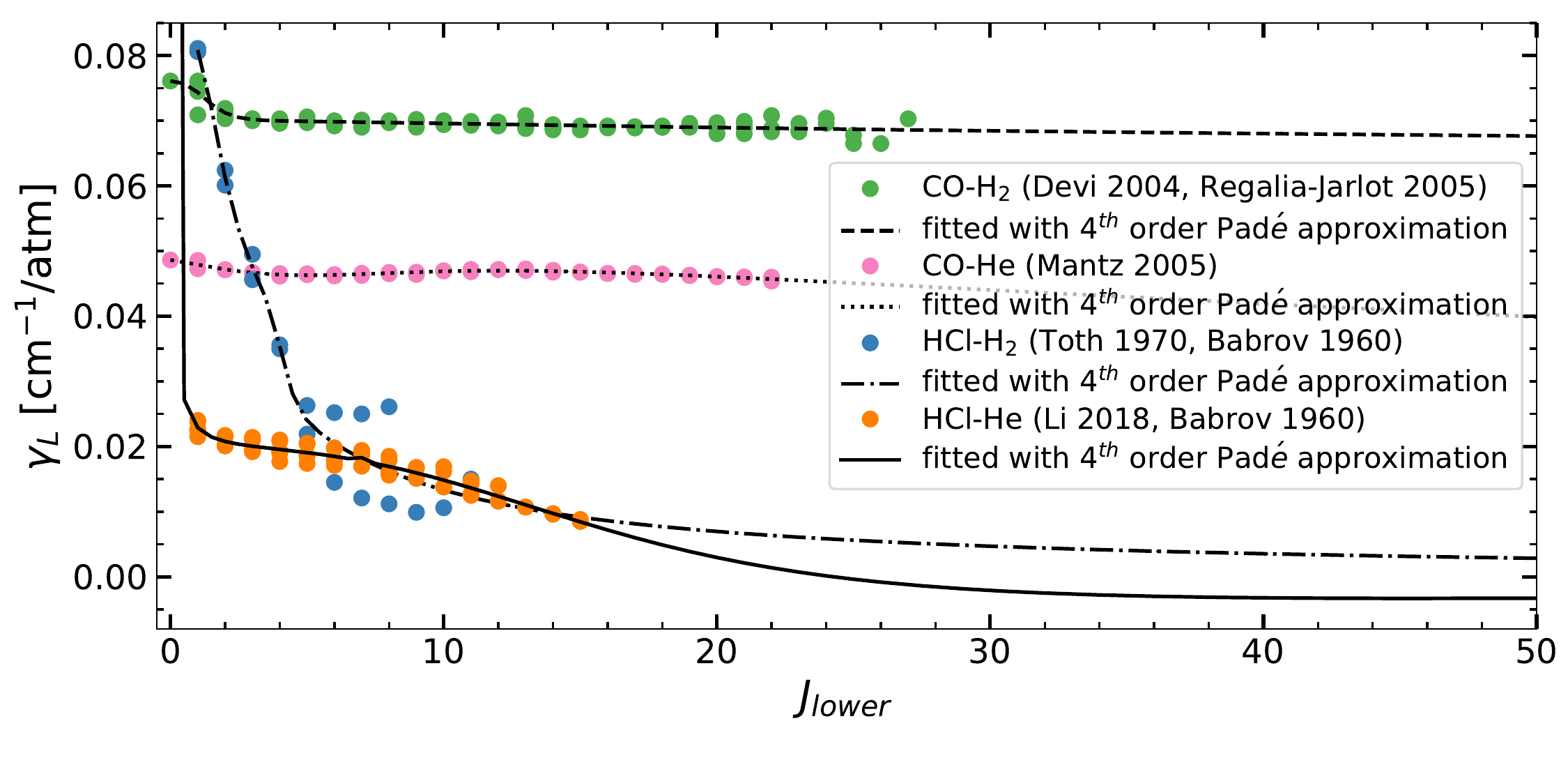}
	  \caption{Comparison of \ce{H2} and He-broadening Lorentz HWHM of CO and HCl molecules using laboratory data. HCl-\ce{H2} system has much stronger $J$-dependency than CO-\ce{H2}, which could be due to their difference in their dipole moments. In contrast, the $J$-dependency of He is lower in both cases. The black lines are also showing the fitted data with 4$^{th}$ order Pad\'e equation (Eq.~\ref{eq:pade-fitting}). Since this equation could go to negative values in some systems and a very large number for $J$=0 (such as HCl-He case), we applied additional limitation (see Eq.~\ref{eq:pade-fitting-constrains}). 
	  We used Eq.~\ref{eq:pade-fitting} to calculate the $J$-dependency of AlH, MgH, CaH, SiO, TiH, FeH, and CrH (see Fig.~\ref{fig:Pbro_fitted-AlH-MgH-etc}). Data of CO and HCl systems in this plot is collected from  \citep{Devi2004_CO-H2-self,REGALIAJARLOT2005-CO-H2,mantz2005-CO-He,Toth1970Hcl-H2He,li2018-HCl-He,babrov1960-HCl-H2He} (see \S\ref{sec:fitting-P-broadening} for more details.).}
	  \label{fig:Pbro_CO_HCl}
\end{figure*}

 {The calculated or estimated Lorentz coefficients, $ \gamma_{\mathrm{L}}(J)$ depend on $J$ quantum numbers. To understand the behavior of $ \gamma_\mathrm{L}(J)$ for all $J$s, we need to consider two issues: 1) the degree or rate of the $ \gamma_{\mathrm{L}}(J)$ decrease with $J$ (or $\frac{\partial\gamma}{\partial J}$), and 2) an equation that can extrapolate these $J$-dependence behavior without becoming negative or increasing sharply.

The first issue can be tackled by comparing the $J$-dependence behavior of the available $\gamma_\mathrm{L}(J)$ coefficients of diatomic molecules with \ce{H2} and He broadening. For example, the Lorentz coefficient of CO molecule (with low dipole moment) has a smooth decrease with high $J$s \citep{Devi2004_CO-H2-self,REGALIAJARLOT2005-CO-H2,mantz2005-CO-He}. In contrast, for absorbers with large dipole moments such as HCl--\ce{H2} and HCl--He systems, $ \gamma_\mathrm{L}(J)$ value drop noticeably according to the laboratory measurements done by \citet{Toth1970Hcl-H2He}, \citet{li2018-HCl-He}, \citet{babrov1960-HCl-H2He} (see Fig.~\ref{fig:Pbro_CO_HCl}).  All the metal hydrides and oxides in question (except AlH) have large dipole moments, and for the lack of any other information we assume that their $J$-dependency may decrease similar to HCl. However, one should keep in mind that dipole moment values are not major factors in the rotational dependence of the broadening parameters. Detailed discussion on the reason for $J$-dependence of spectral lines is provided by \citet{hartmann2021Textbook} (chapter 3) and \citet{Renaud2018H2OLineShape}.
The second issue may be solved by using the proposed equation by the HITRAN group\footnote{\href{www.hitran.org}{https://www.hitran.org}} \citep[e.g., see][and their upcoming paper, Gordon et. al. 2021]{Gordon2017HITRAN2016}. As part of their group efforts, \citet{Tan2019H2O-Pbro} have examined the fitting accuracy of the available pressure-broadening coefficients for water as a broadener and several terrestrial key absorbers (e.g., CO, \ce{CO2}, \ce{CH4}); and they suggested the fourth‐order Pad$\acute{e}$ approximant as the most reliable prediction of the Lorentz $J$-dependency. Fourth‐order Pad$\acute{e}$ equation provides a smooth decrease of $\gamma_\mathrm{L}$ with $J$ and has the following form:

 \begin{equation}\label{eq:pade-fitting}
 \gamma_\mathrm{L}(J)= \frac{a_0 + a_1 J + a_2 J^2 +  a_3 J^3}{1 + b_1 J + b_2 J^2 +  b_3 J^3+  b_4 J^4}
 \end{equation}
 
 where $a_0$ to $a_3$ and $b_1$ to $b_4$ are the fitting coefficients. $J$ is also the lower total angular quantum number. Besides, we applied the following constraints to this equation to avoid negative numbers of $\gamma_\mathrm{L}$ as well as follow a sharp $J$-dependency for absorbers with large dipole moments (i.e., MgH, CaH, CrH, FeH, TiH):

\begin{equation}\label{eq:pade-fitting-constrains}
\begin{aligned}
& \gamma_\mathrm{L}(J=10)  = \sim0.5\gamma_\mathrm{L}(J=0) \\
& \gamma_\mathrm{L}(J=40)  = \sim0.1\gamma_\mathrm{L}(J=0) \\
& \mathrm{if} \  \gamma_\mathrm{L}(J > 40) < 0  => \gamma_\mathrm{L}(J)=\gamma_\mathrm{L}^{min} - (\frac{1}{J}*\gamma_\mathrm{L}^{min})\\
\end{aligned}
\end{equation}}

\begin{figure*}[htb!]
\centering
\includegraphics[width=\textwidth]{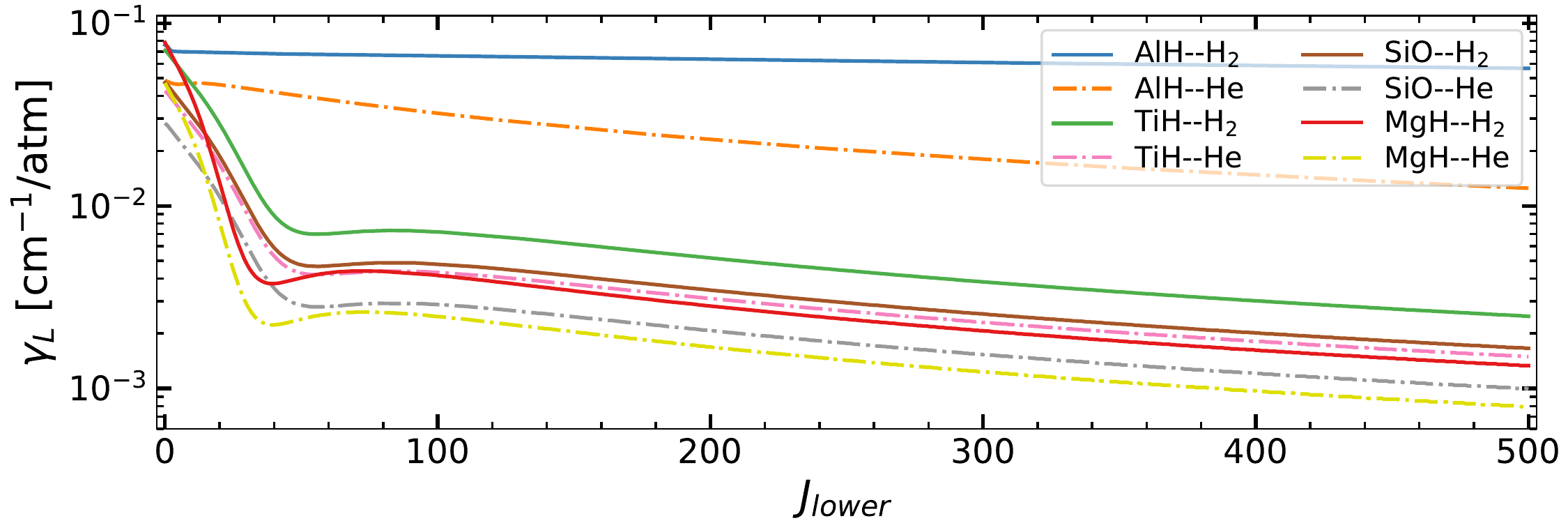} 
	  \caption{Example of used Lorentz coefficients for AlH, TiH, SiO, and MgH. The AlH Lorentz coefficients are assumed to be equal to CO, because they both have the same dipole moments. For other absorbers, the Lorentz coefficient are decreasing sharply up to $J$=40 similar to HCl because they all have high dipole moments. 4$^{th}$ order Pad\'e equation (Eq.~\ref{eq:pade-fitting} including the constrains introduced in (Eq.~\ref{eq:pade-fitting-constrains}) is used and the coefficients are reported in Table~\ref{tab:fitting-coefficients-Pade} in order to reproduce these broadening data (see \S\ref{sec:fitting-P-broadening} for more details.).}
	  \label{fig:Pbro_fitted-AlH-MgH-etc} 
\end{figure*}

An example of the calculated broadening coefficients for the metal hydrides/oxides in question is presented in Fig.~\ref{fig:Pbro_fitted-AlH-MgH-etc}. Additionally,, the Fourth‐order Pad\`e coefficients for all the absorbers in this study are listed in Table \ref{tab:fitting-coefficients-Pade} and are valid up to $J$=500.

\subsection{Wing cut-off}\label{sec:wing Cut-off} 
  Wing cut-off is an important challenge in fully and accurately computing the opacity continuum (see \S\ref{sec:opacity-challenge}). Given the lack of sufficient pressure broadening data for these species and the huge computational time, the Voigt profile is implemented in various pre-computed opacity data. The current method to deal with this non-Lorentzian behavior of the Voigt profile is either using absolute wavenumber such as 100 cm$^{-1}$ \citep[e.g.,][]{macdonald2019POSIEDON} or just using multiplication of the Voigt profile (e.g., 200 to 500 of Voigt HWHM) \citep[e.g.,][]{Chubb2020,Freedman2014}. 
 
 Note that several laboratory and theoretical studies have shown that non-Voigt behavior provides a more accurate opacity continuum \citep[e.g.,][]{Ngo2012H2O_nonVoigt-profile}. For example, \citet{Hartmann2002wing} have recorded the 3.3~$\micron$ spectrum of \ce{CH4} in \ce{H2} bath gas and have shown that the Voigt behavior is valid up to 25 cm$^{-1}$ from the line center. Inspiring by their results, we decide to use absolute number for the wing cut-off as follow: 30 cm$^{-1}$ for $P\leq200$~bar and 150 cm$^{-1}$ for $P > 200$~bar. We noticed that \citet{molliere2019petitradtrans} have considered this method as well in calculating their opacities.
 
 It is worth mentioning that we did not use the multiplication factor of the Voigt linewidth because for a generic spectral line at low-to-moderate pressure and at the infrared spectral region with small wavenumber (such as $P<$10$^{-3}$~bar at $<$1000 cm$^{-1}$), the 500$\times$Voigt-half-width will be less than 10cm$^{-1}$ which is not sufficiently high value for the wing cut-off according to the results by \citet{Hartmann2002wing}. Additionally, for high pressures (such as $P>$100bar), this 500$\times$Voigt-HWHM is a large number even larger than 500 cm$^{-1}$ which results in an extreme overestimation.

\subsection{Line List}\label{sec:linelist} 

A typical line-list data contains all of the necessary ingredients to generate ACSs, such as line positions, intensities, rotational and vibrational quantum assignments, electronic assignments, and also various pressure-broadening and pressure-shift values where available. The completeness and accuracy of line-lists are a key focus of both laboratory and {\it ab initio} studies. In this work, we use the latest version of line list data provided by the ExoMol group \citep{Tennyson2020Review}, summarized in  Table~\ref{tab:Summary-linelist}.

\subsection{Computational Details}\label{sec:computational-details-ACS}

We employ the {\tt  ExoCross}\footnote{\href{https://github.com/Trovemaster/exocross/wiki}{github.com/Trovemaster/exocross/wiki}} code developed by \citet{yurchenko2018exocross} to compute the temperature-dependent line strengths and the Voigt profile \citep{Humlicek1979} for every individual line under the conditions given in Table \ref{tab:comp}. We assume H/He broadening governed by background atmospheres composed of 85\% \ce{H2} and 15\% He--typical Jovian-like atmosphere conditions. The spectral sampling resolution is optimized as a function of temperature, pressure, and spectral subdivisions in such a way as to fully resolve the individual lines.

\section{Results: Absorption Cross-Section }\label{sec:Results}

In this section, we present a brief discussion on the line-list source for each absorber, as well as illustrative examples of the pre-computed ACS data. Then, we compare our generated ACS data with the currently available database's.

 \subsection{H$_2$O}

\begin{figure*}[!htb]
\centering
\includegraphics[width=\textwidth]{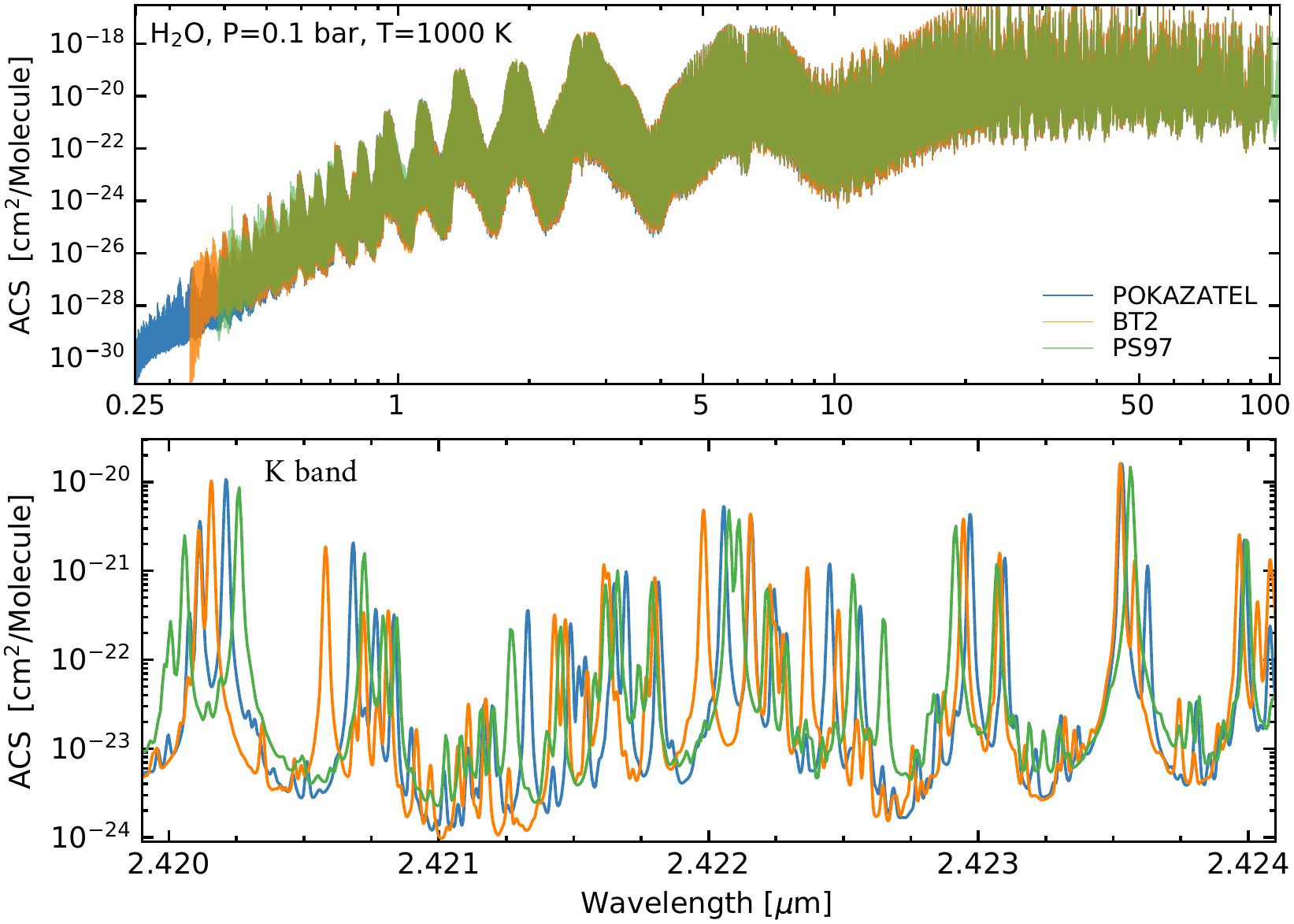} 
	  \caption{Comparison of different \ce{H2O} ACS for 0.1 bar and 1000K, including  POKAZATEL, BT2, and PS97. Note, that $POKAZATEL$ and $BT2$ line lists are from ExoMol group, however, $POKAZATEL$ is the most recent version. In addition, the Schwenke ACS is generated by \citet{Freedman2014} using \citet{Partridge1997} water line list (named $PS97$). Their expanded comparisons for 2.4~$\micron$ (bottom)  shows some differences in their line positions, intensities. Additionally, there are some extra lines from POKAZATEL that are not exist in others. Although they are weak, but they can change the opacity continuum, and so their inclusion is important in particularly windows spectral regions (see \ref{sec:opacity-challenge} for more discussion). These differences result from the implemented quantum mechanic level of complexity, and they could bias the interpretation of medium-to-high-resolution observed data.}
	  \label{fig:H2O-compareXS-DifferentLinelist} 
\end{figure*}

\cite{Partridge1997} used high-level {\it ab initio} approach to computing the potential energy surface and dipole moment function of water. They then empirically adjusted their water line list (hereafter, $PS97$) by comparing their line position with available laboratory spectra. About a decade later, the ExoMol team \citep{Barber2006} computed the $BT2$ water line list, which includes $\sim$twice as many transitions with a somewhat lower total angular quantum number $J$ but over a wider spectral range, extending deeper into the blue-optical. However, the comparison of both line lists with laboratory \ce{H2O} measurements showed that the $PS97$ ACSs are more accurate in predicting line positions than $BT2$ at wavelengths longer than 1~$\micron$  \citep{Alberti2015Water-validation,Melin2016Water-validation}. It is for this reason that the $PS97$ list has been used in several works (e.g., \cite{Freedman2014, Marley2015}), despite more modern line lists. Meanwhile, HITEMP---the high-temperature arm of the CfA HITRAN database---was created, using a trimmed version of the $BT2$ line list, but with rigorous validation and adjustments against laboratory data \citep{Rothman2010}. Recently, ExoMol  improved the $BT2$ line list by refining the potential energy surfaces, and extending the wavelength range down to 0.25~$\micron$ ~(40,000 cm$^{-1}$), referred to as the $POKAZATEL$ \citep{Polyansky2018-Water-POKAZATEL} line list. Table~\ref{tab:H2O-linelists} summarizes these line lists and their key differences, and Fig.~\ref{fig:H2O-compareXS-DifferentLinelist} compares the resultant cross-sections for a representative pressure/temperature. Besides the spectral range, there are extensive differences in their line positions, line shifts with pressure, and the number of lines, clearly apparent in different spectral regions such as the 2.420--2.424~$\micron$ band (K band)--subject to numerous high-resolution observations (e.g., \cite{Birkby2013water-highRes}). \citet{Brogi2019} showed that these differences matter in constructing high-resolution spectroscopy cross-correlation functions potentially leading to bias in the interpretation of the atmospheric composition. Recently, \citet{Gandhi2020XS-HighRes} compared the water $POKAZATEL$ data with the laboratory HITEMP data, also in the context of high-resolution cross-correlation spectroscopy, showing very good agreement up to temperatures of $\sim$1200 K.   

\begin{figure}[!htb]
\centering
\includegraphics[scale=0.98]{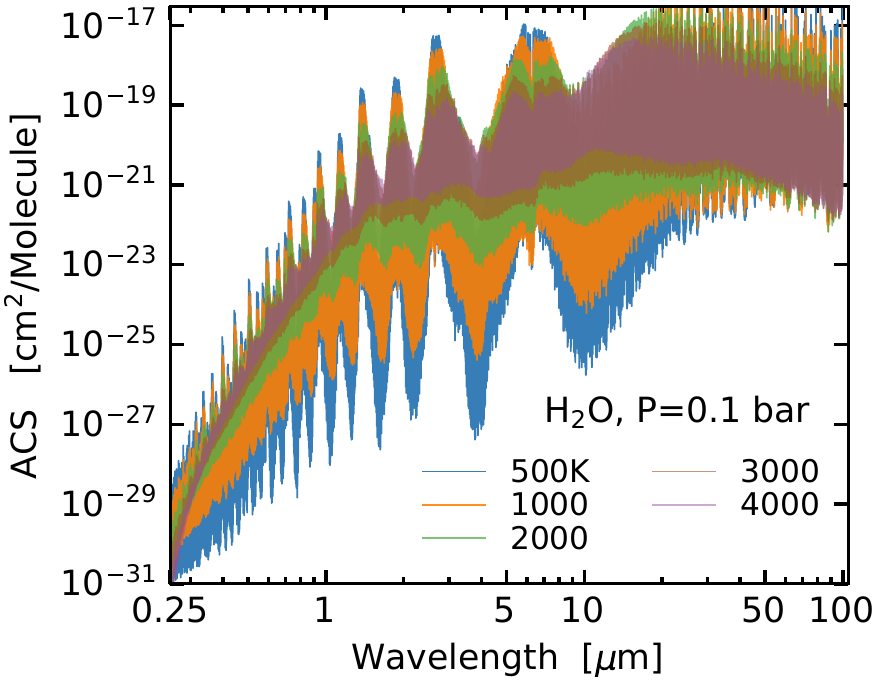} 
	  \caption{Example of \ce{H2O} $POKAZATEL$ ACS spectra for 0.1 bar and multiple temperatures. For the same pressure, ACS with higher temperature have broader Doppler linewidth, and this will results in larger opacity in the line core comparing to the wings.}
	  \label{fig:H2O} 
\end{figure}

 In our ACS database, we used the $POKAZATEL$ \ce{H2O} line list over the full spectral range of 0.25--100~$\micron$ at the temperature and pressure conditions given in Table \ref{tab:comp}. Figure~\ref{fig:H2O} shows a subset of our ACSs as a function of temperature.

{More recently,  \citet{Conway2020} have published their \ce{H2O} line list calculated from experimental spectroscopic data, which includes $\sim$10$^{6}$ transitions and it is limited up to $J_{max}$=20. Their line lists have shown a significant improvement in the line position for the ultraviolet spectral region. Note that the current ground-based telescopes with very high resolution are mostly detecting in the near-IR region; however, future telescopes in the UV-Vis regions are required to have accurate line positions for high-resolution cross-correlation modeling of observational data.}

\subsection{TiO}\label{sec:TiO}

Several spectroscopic measurements have been carried out to determine the spectroscopic parameters such as electronic dipole moments, hyperfine spectral properties, and rotational and vibrational constants for the TiO electronic ground state\footnote{In diatomic spectroscopy, in general, the electronic ground state is represented by $X$ and the higher electronic states are labeled by $A$, $B$, and so on. In addition, the electronic multiplicity (2S+1) and the total orbital angular momentum ($\Lambda$) are used to represent the electronic state. For example, $X^3\Delta$ is the ground state of TiO with a multiplicity of 3 and orbital angular momentum of $\Delta$, which is equivalent to $\Lambda$=2. Hence, $C^3\Delta-X^3\Delta$ is showing the electronic transition between the TiO state $C$ and $X$, with the same total orbital and spin angular momentums. Chapter 9 by \citet{Bernath_book} provided detailed information on this topic.}, $X^3\Delta$, as well as the low-lying excited states such as $B^3\Pi$, $C^3\Delta$, and $D^3\Pi$ \citep{Steimle2003,Amiot1995,Balfour1993}. In addition, several laboratory measurements have been done to determine the line position and strength between the following TiO electronic states: ($C^3\Delta-X^3\Delta$) \citep{Hodges2018}, ($c^1\Phi-a^1\Delta$) \citep{Bittner2018},
($b^1\Pi-a^1\Delta$) \citep{Ram1996TiO,Bittner2018}, ($A^3\Pi-X^3\Delta$) \citep{Ram1999}, ($f^1\Delta-a^1\Delta$) \citep{Brandes1985}, and ($c^1\Phi-a^1\Delta$) \citep{Bittner2018}.

\citet{Kurucz1992OpacityData} collected the available laboratory spectroscopic data for all  TiO transitions to make the first comprehensive laboratory TiO line list (hereafter TiO-$Kurucz92$).  Concurrently, \citet{Plez1992} used a theoretical spectroscopic method to calculate the transition dipole moments to generate TiO opacities. A few years later, they used more accurate dipole moments from \citet{Langhoff1997} as well as the available spectroscopic measured data to improve line positions and intensities of the TiO nine low-lying electronic states \citep{Plez1998}.

\citet{schwenke1998} independently used {\it ab initio} methods that included the spin-orbit and rotation-orbit coupling to compute potential energy surfaces for 13 electronic states. These surfaces were used as a Rydberg–Klein–Rees (RKR) potential to extract rovibrational energy levels. \citet{schwenke1998} also adopted the transition moments from \citet{Langhoff1997} to determine the intensities for each rovibrational transition in order to compute a complete line list (intensities and positions) for 45$\times$10$^6$ lines. \citet{Allard2000} compared these two ($Plez92$ and $Schwenke98$) TiO line lists in the context of M- and Brown-dwarf atmospheres, finding that the $Schwenke98$ data could better model the thermal emission from cool substellar atmospheres.

\begin{figure*}[!htb]
\centering
\includegraphics[scale=0.8]{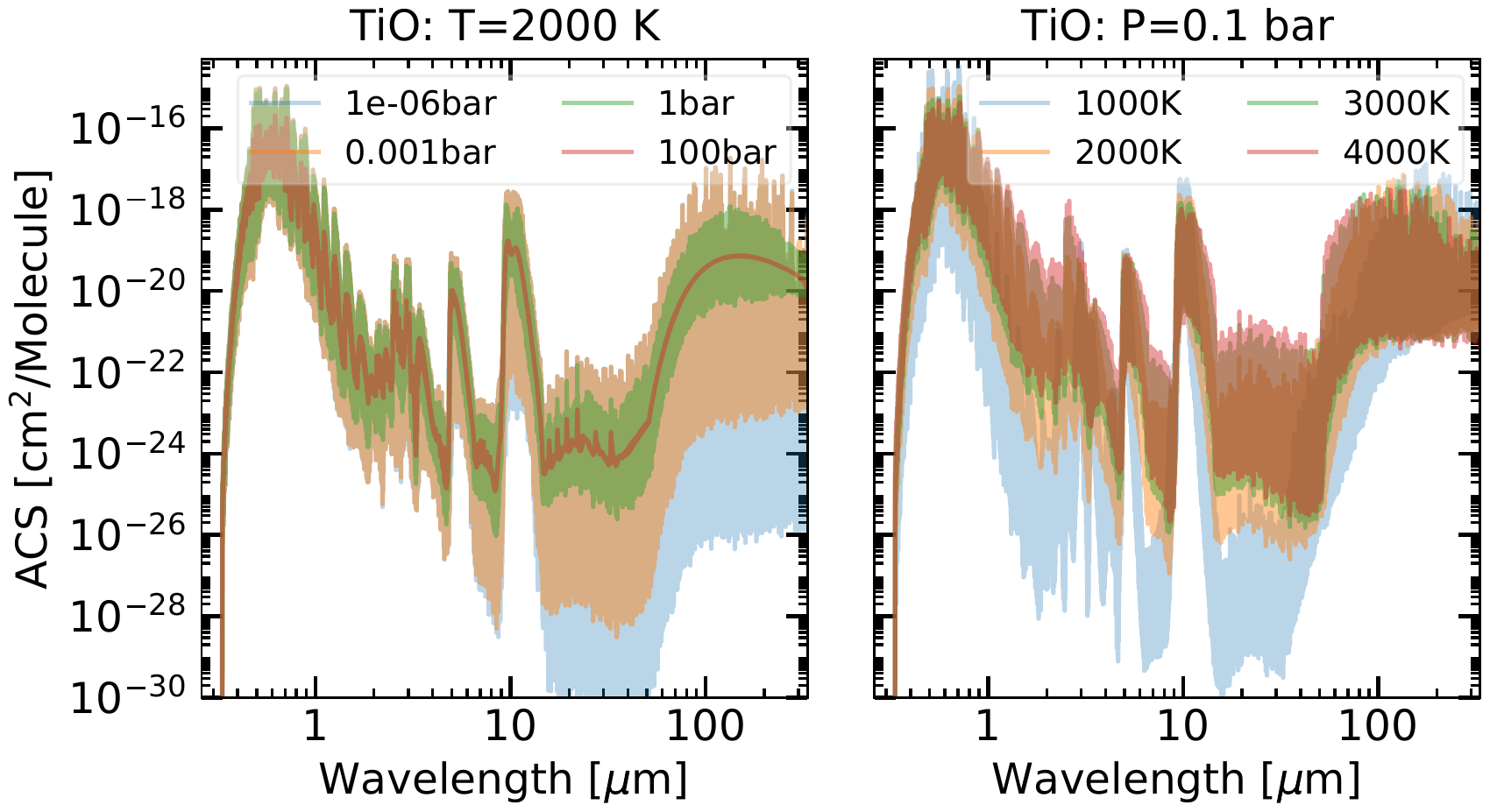} 
	  \caption{Example of the pre-generated TiO-$TOTO$ ACS data: at a constant pressure, Doppler effect is the main source of line-width, particularly at low wavelengths (right). In contrast, at constant temperature, the higher pressure shows the most broadening as a result of collisional effect (left). Note that Lorentz width is dependent on the temperature as well. We used \citep{McKemmish2019TiO-TOTO} line list to  generated these ACS  data.}
	  \label{fig:TiO-TOT} 
\end{figure*}

\begin{figure*}[!htb]
\centering
\includegraphics[scale=1.0]{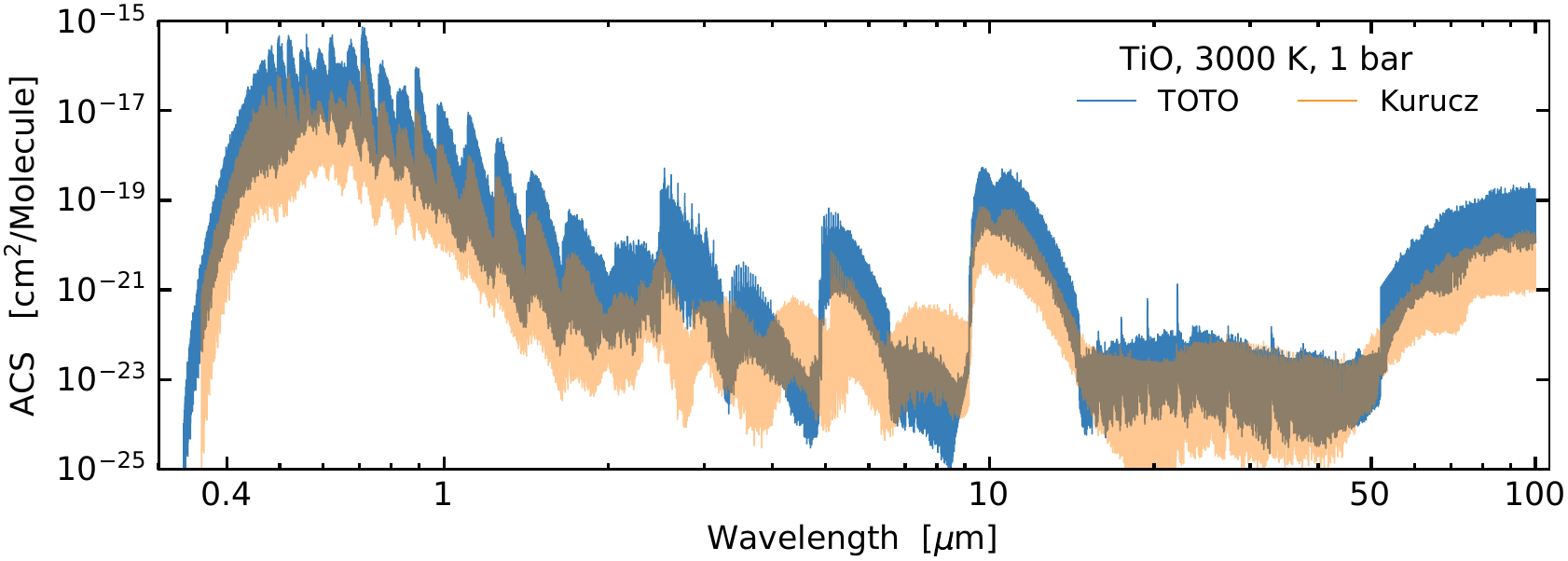} 
	  \caption{Compare TiO absorption cross-sections generated from {\it TOTO} and {\it Kurucz} line list for 1 bar and 3000 K. Note that there are noticeable differences in the whole spectral range between both cases.  2.5--9 $\mu m$ region is showing the highest inconsistency between two spectra. { Overall, the TiO $TOTO$-ACS is showing a larger value than the Kurucz data, which might be due to its large number of transitions as well as the modified electronic states which results in shift in some spectral bands.}}
	  \label{fig:TiO_TOTO_Curuz} 
\end{figure*}

\begin{figure*}[!htb]
\centering
\includegraphics[scale=0.8]{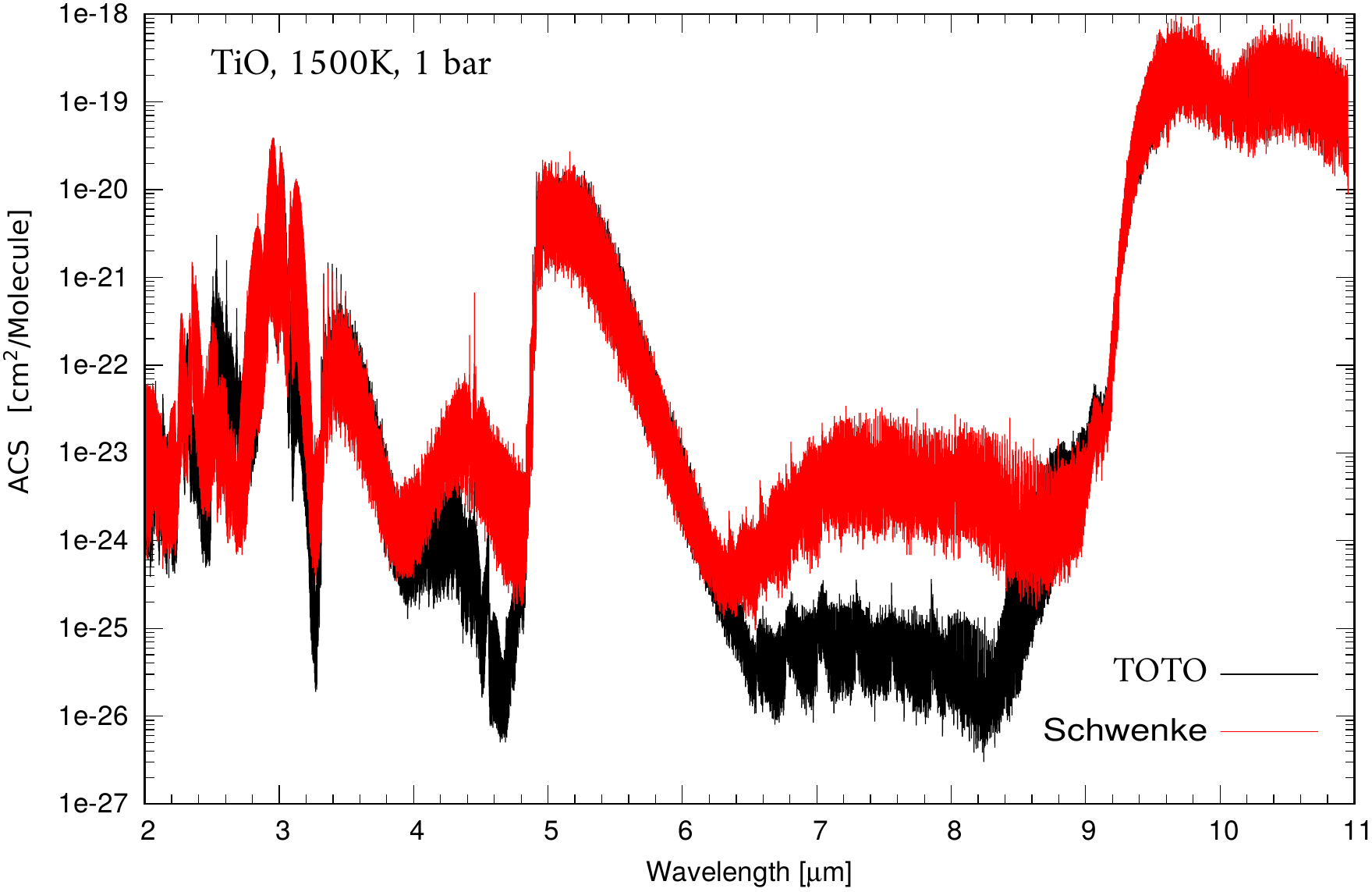} 
	  \caption{Compare TiO absorption cross-sections generated from the $TOTO$ and $Schwenke$ line list for 1 bar and 1500 K. There are noticeable differences in the whole spectral range between both cases. The highest inconsistency is at spectral regions $\sim$4--5 and 6.5--8.5~$\micron$, that results in a dramatic bias in modeling low and high resolution synthetic emission or transmission spectra. One possible explanation would be the difference between the number of rotational levels that have been accounted by both  $TOTO$ and $Schwenke$ linelist. }
	  \label{fig:TiO_TOTO_Schwenke} 
\end{figure*}

\citet{Hoeijmakers2015} used $Schwenke98$ and $Plez92$ line lists to detect/identify TiO in the HD 209458b atmosphere via high-resolution cross-correlation spectroscopy (R$\approx$100,000) finding that the line positions in both lists are not sufficient to extract an accurate signal. These mounting inadequacies in line positions and intensities motivated the need for a better, more comprehensive, TiO line list. ExoMol used a high-quantum level of basis set computation methods with spin-orbit coupling to generate all rovibronic (rotation-vibration-electronic) transitions between the 13 low-lying electronic states, covering a spectral range of 0.33--100~$\micron$ (called $TOTO$, \cite{McKemmish2019}).  In our ACS database we used this $TOTO$ line list\footnote{ \href{http://exomol.com/data/molecules/TiO/49Ti-16O/Toto/}{exomol.com/data/molecules/TiO/49Ti-16O/Toto/}} from 0.3-100~$\micron$ over the conditions given in Table \ref{tab:comp}.  Figure~\ref{fig:TiO-TOT} summarizes a subset of these ACSs and Figure~\ref{fig:TiO_TOTO_Curuz} and \ref{fig:TiO_TOTO_Schwenke} compare to past line lists using identical computational setups.

More recently, \citet{Bernath2020TiO} have recorded the TiO spectra in the visible--near-infrared regions (476–1176 nm) at 2300 K, and compared the measured intensity and line position with the TOTO data. In addition, \citet{Nugroho2020KELT20b} applied high-resolution cross-correlation spectroscopy method using the TiO $Plez1998$ (updated version) and $TOTO$ line lists, but they could not detect TiO signature in KELT-20b atmosphere. Note that even though both \citet{Nugroho2020KELT20b} and our study have used the same TiO-$TOTO$ line list, our generated ACS data are different from their data because of the following reasons. First, their resolution is 100 points/wavenumber, however, our resolution is at least 1 times finer than them for several pressure-temperature points. Second, Their line wing cut-off is 100~cm$^{-1}$ which is different from ours. Besides, our pre-generated ACS data covers a wider temperature and pressure range with larger grid points than the current pre-generated ACSs.

\subsection{VO}\label{sec:VO}

\begin{figure*}[!htb]
\centering
\includegraphics[scale=1.0]{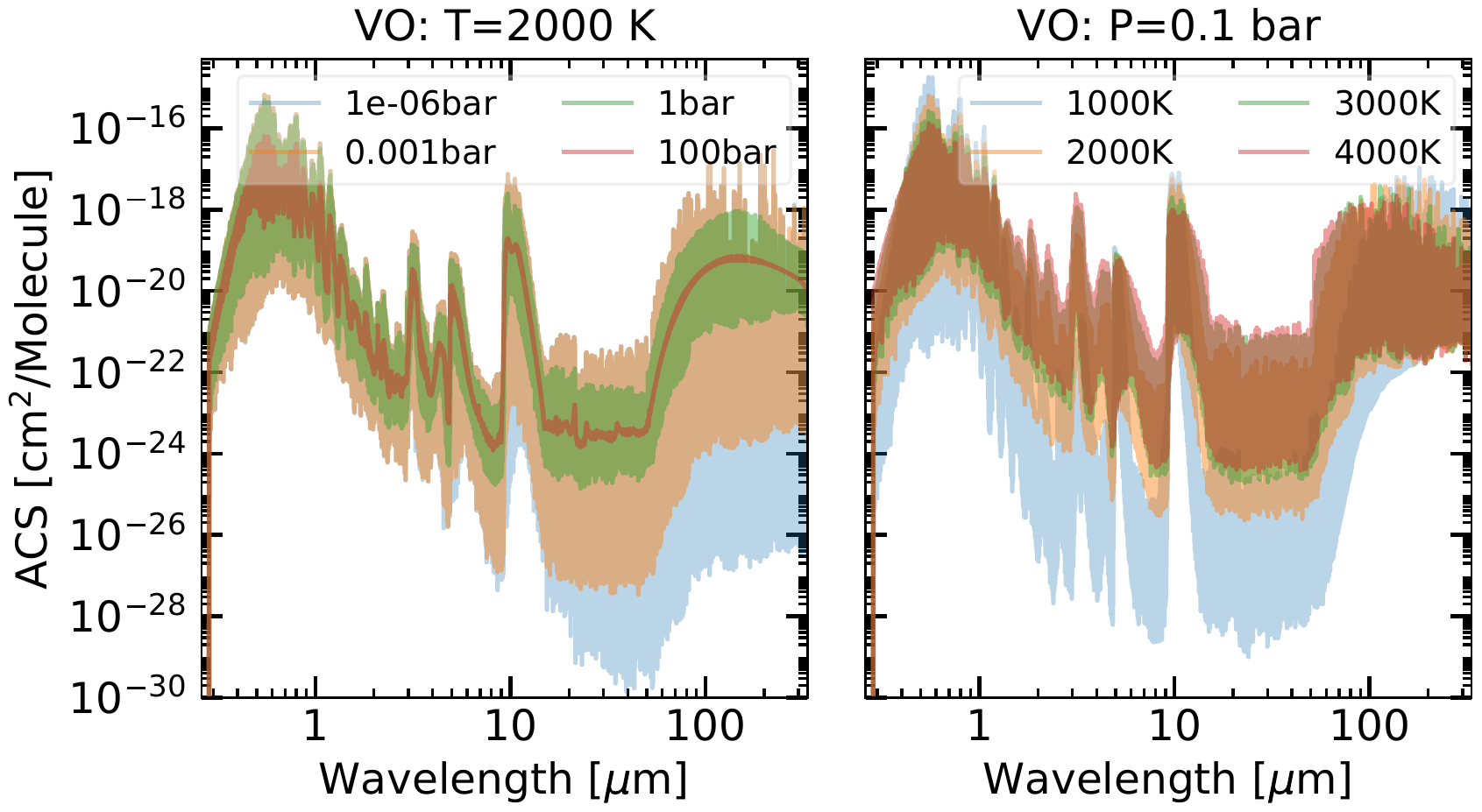} 
	  \caption{Example of our pre-generated VO-VOMYT absorption cross-sections: at a constant pressure, Doppler effect is the main reason of controlling the line-width, particularly at low wavelengths (right). In contrast, at constant temperature, the higher pressure shows the most broadening as a result of collisional effect (left). These pre-generated ACSs are from \citep{McKemmish2016} line list data, which consists of the 13 low-lying electronic state and have $\sim$227 million lines.  }
	  \label{fig:VO} 
\end{figure*}

\begin{figure*}[!htb]
\centering
\includegraphics[scale=0.8]{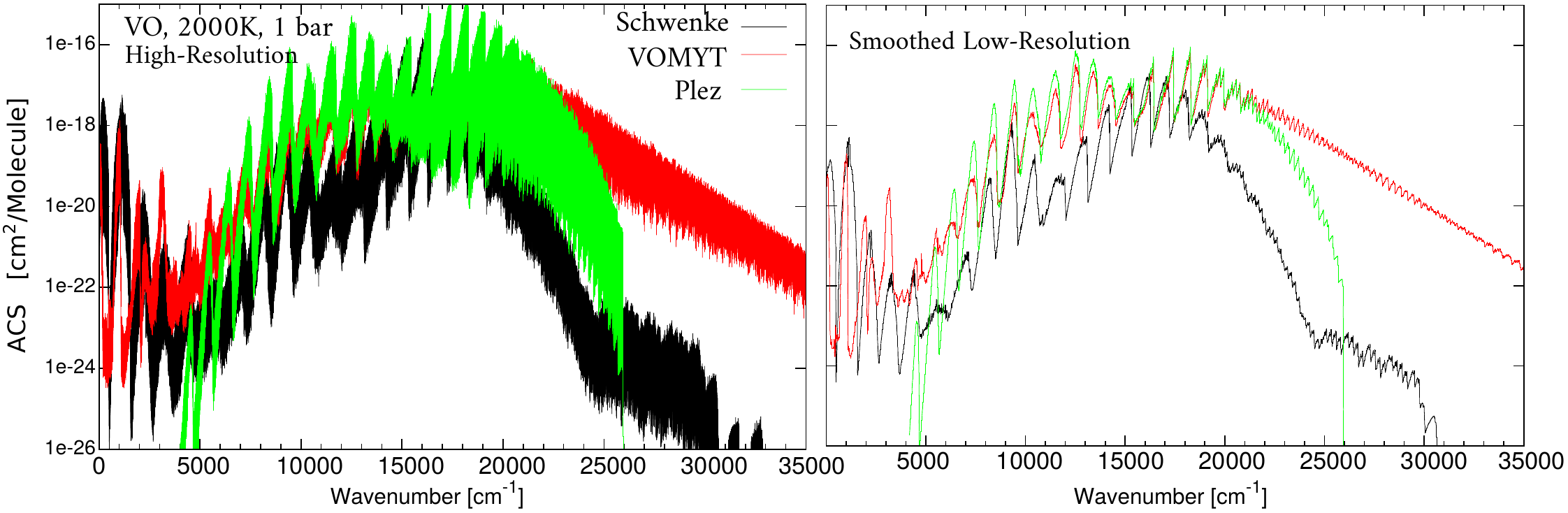} 
	  \caption{Comparison of different VO absorption cross-sections generated from different line lists: $Schwenke$, $VOMYT$, and $Plez$ at 2000~K and 1~bar. $VOMYT$ line list \citep{McKemmish2016} includes the 13 low-lying states and $\sim$227 million transitions, and hence the intensity of its spectra (red) is higher than other line lists (except $Plez$ around 10000 cm$^{-1}$ which resulted from A--X transitions). In contrast, $Plez$ line list \citep{Plez1999-VO}(green) has only the  A--X, B--X, and C--X bands with no infrared transitions, and so it is only showing data at 5000-25000 cm$^{-1}$(0.4--2 ~$\micron$ ). The central message of this figure is to show the difference between VO-ACS data from different line lists is remarkable, and naive implementation of VO ACS data could result in false conclusions and misinterpretation of the observational exoplanet/brown-dwarf spectra.}
	  \label{fig:VO-compare-VOMYT-Plez} 
\end{figure*}

Numerous laboratory works have focused on determining the spectroscopic constants of the ground-state ($X^4\Sigma^{-}$) and low-lying excited states, including the $A-X$ ($\sim$1.1~$\micron$ ~ \citep{CHEUNG1982}), $B-X$ ( $\sim$0.74~ $\micron$ ~ \citep{HUANG1992}), and $C-X$ ($\sim$0.43--0.58~$\micron$ ~ \citep{Hopkins2009}) systems as well as quartet–quartet transitions (e.g., $^4\Delta-^4\Pi$ and $^4\Delta-^4\Phi$, $\sim$0.85 and 1.05~$\micron$ ~\citep{Merer1987}) and doublet--doublet transitions (e.g., $^2\Delta-^2\Delta$ and $^2\Phi-^2\Delta$, $\sim$0.45 and 1.4~$\micron$ ~\citep{RAM2002,Ram2005VO}.
Complimenting the experiments, several theoretical {\it ab initio} works on VO have been undertaken to assign all rovibronic transitions to improve the atmospheric modeling of ultra-hot atmospheres (e.g., \citep{Miliordos2007}). \citet{Hubner2015} computed potential energy surfaces of the electronic ground and excited states and their corresponding spectroscopic constants (equilibrium radius, dissociation energies). \citet{McKemmish2016} (hereafter, $VOMYT$) improved upon these previous {\it ab initio} results by including spin-orbit and spin-spin couplings as well as by leveraging available spectroscopic data to modify the potential energy surface to improve line positions and intensities, resulting in a robust VO line list with 2.77$\times$10$^8$ transitions.
We used this $VOMYT$ line list\footnote{ \href{http://exomol.com/data/molecules/VO/51V-16O/VOMYT/}{exomol.com/data/molecules/VO/51V-16O/VOMYT/}} to generate our VO cross-sections given the parameters in Table \ref{tab:comp}. A sample of these cross-sections is shown for multiple temperatures and pressures in Fig.~\ref{fig:VO}. { In addition, Fig.~\ref{fig:VO-compare-VOMYT-Plez} illustrates the difference between the $VOMYT$ ACS data with the old ones computed from the $Plez$ and $Schwenke$ line lists for the full UV-Vis-IR range. The reason for these considerable differences is due to the calculated number of transitions and the computed electronic states from each line list. For example, \citet{Plez1999-VO} has accounted for 3.1 million lines, while $VOMYT$ has 277 million transitions, which explains why it has a higher ACS value. The number of transitions in the $Schwenke$ line list is much smaller than $VOMYT$ as shown in the figure. Further details can be found in table 19 by \citet{McKemmish2016}.}

\subsection{FeH}\label{sec:FeH}
\begin{figure*}[!htb]
\centering
\includegraphics[scale=0.85]{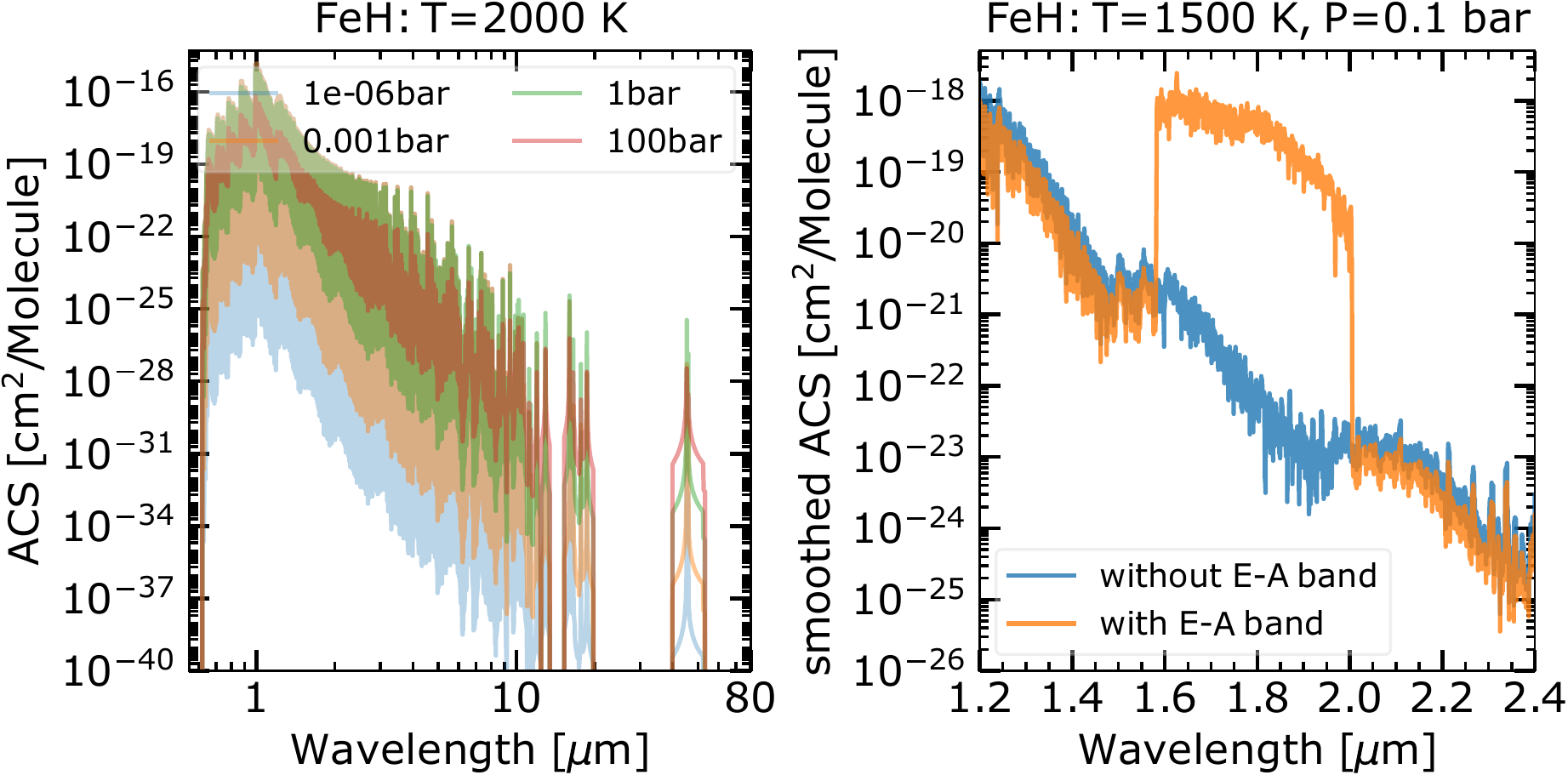} 
	  \caption{(left) An example of our FeH ACS spectra generated from \citet{Dulick2003} linelist for multiple pressures and 2000~K. (right) Comparison of the FeH absorption cross-sections generated with (orange) and without (blue) including the $E-A$ band. The discrepancies in these ACS data are due to the lack of sufficient accuracy in the $E-A$ band.}
	  \label{fig:FeH-XS} 
\end{figure*}

FeH (iron monohydride) is a diatomic radical with a high-multiplicity ground state ($X^4\Delta$), resulting in a complex electronic spectrum (see page~71 in \citet{veillard2012}). Several laboratory spectroscopic measurements have been performed since the 1970s to record and analyze the FeH $F^4\Delta-X^4\Delta$ 0--0 band (known as Wing-Ford band) \citep{McCormack1976FeH} as well as to characterize the infrared FeH spectra \citep{Phillips1987FeH-IR}. Later, \citet{Dulick2003} used an {\it ab initio} quantum mechanic approach to calculate a new line list for several vibrational bands in the $F^4\Delta-X^4\Delta$ system, including the electronic transmission dipole moment and spectroscopic constants (e.g., rotation and vibration constants). Following their theoretical work, \citet{Wende2010} used high-resolution spectroscopy to record the Wing-Ford spectral lines in 989.8--1076.6~nm. In the same year, \citet{Hargreaves2010} used the above laboratory data to investigate the $E^4\Pi-A^4\Pi$ ($\sim$1.58~$\micron$) rovibrational spectra. Note that the band strength of the $E-A$ system calculated from this work has been problematic in several astrophysical spectral observations, further discussed in \S\ref{sec:Comments on FeH line lists}, and so we did not include this $E-A$ band in generating our FeH opacities. More recently, \citep{Bernath2020MOLLIST} synthesized these works into the MoLLIST line database.
We used the FeH line list by \citet{Dulick2003}\footnote{ \href{http://exomol.com/data/molecules/FeH/56Fe-1H/MoLLIST/}{exomol.com/data/molecules/FeH/56Fe-1H/MoLLIST/}; Note that the reference given in the ExoMol website is mistakenly to the recent work by \citet{Wende2010}, but the provided line lists there are from \citet{Dulick2003}.} to compute our cross-sections over the properties given in Table \ref{tab:comp}. Fig.~\ref{fig:FeH-XS} (left) illustrates a few pressure points from our grid and Fig.~\ref{fig:FeH-XS} (right) shows the results of including the \citet{Hargreaves2010} $E-A$ band.

\subsubsection{Comments on FeH Line Lists}\label{sec:Comments on FeH line lists}
 The history of the FeH line lists is complex and deserves an expanded discussion. Much of the uncertainty involves the $E^4\Pi-A^4\Pi$ electronic transition near 1.58~$\micron$ (6300 cm$^{-1}$, Fig.~\ref{fig:FeH-XS} (right panel). \citet{Hargreaves2010} obtained the positions and intensities within this band using a spectrum previously recorded with a Fourier Transform Spectrometer at Kitt Peak National Observatory from an emission spectrum produced by a King-type furnace as the source \citep{Phillips1987FeH}. A calibration step was necessary to convert the measured emission intensities, without an absolute intensity scale, into line strengths at a given temperature, resulting in their line list for this $E-A$ transition.

However, there are several potential sources of uncertainty in this measured intensities-to-line strengths conversion. First, the temperature of the tube furnace was determined to be closer to 2200 K based on the relative line intensities (instead of the 2673 K as recorded by \citet{Phillips1987FeH}). The actual final calibration connecting the lab data to intensities was done using lines that had been individually identified by \citet{Balfour2004FeH}. The spread in the intensities (see Figure 1 in \citet{Hargreaves2010} paper) is quite large and a linear regression was used to obtain a conversion factor for the intensities. \citet{Hargreaves2010} mentioned in the paper that points used to calibrate the laboratory intensities contained considerable scatter, and the change in the assumed temperature may play a role in the intensity problem. In addition, An average lower state energy (2250 cm$^{-1}$) is used for all unassigned lines (96$\%$ of the lines provided), which would contribute to the uncertainty at temperatures away from 2200 K as the line list from this work was provided at 2200~K. Two points to note: the ratio of the partition function $Q$(2673K)/$Q$(2200K) = 1.64 and the change in the factor needed to convert emission intensities to absorption values (see Eq.~6 in \cite{Hargreaves2010}) is approximately
a factor of 2 when considering the ratio of the two temperatures. These caveats may play a role in astrophysical data-model comparisons, suggesting that this FeH band, in particular, deserves further theoretical and laboratory attention.

\subsection{TiH}\label{sec:TiH}

\begin{figure*}[!htb]
\centering
\includegraphics[scale=1.0]{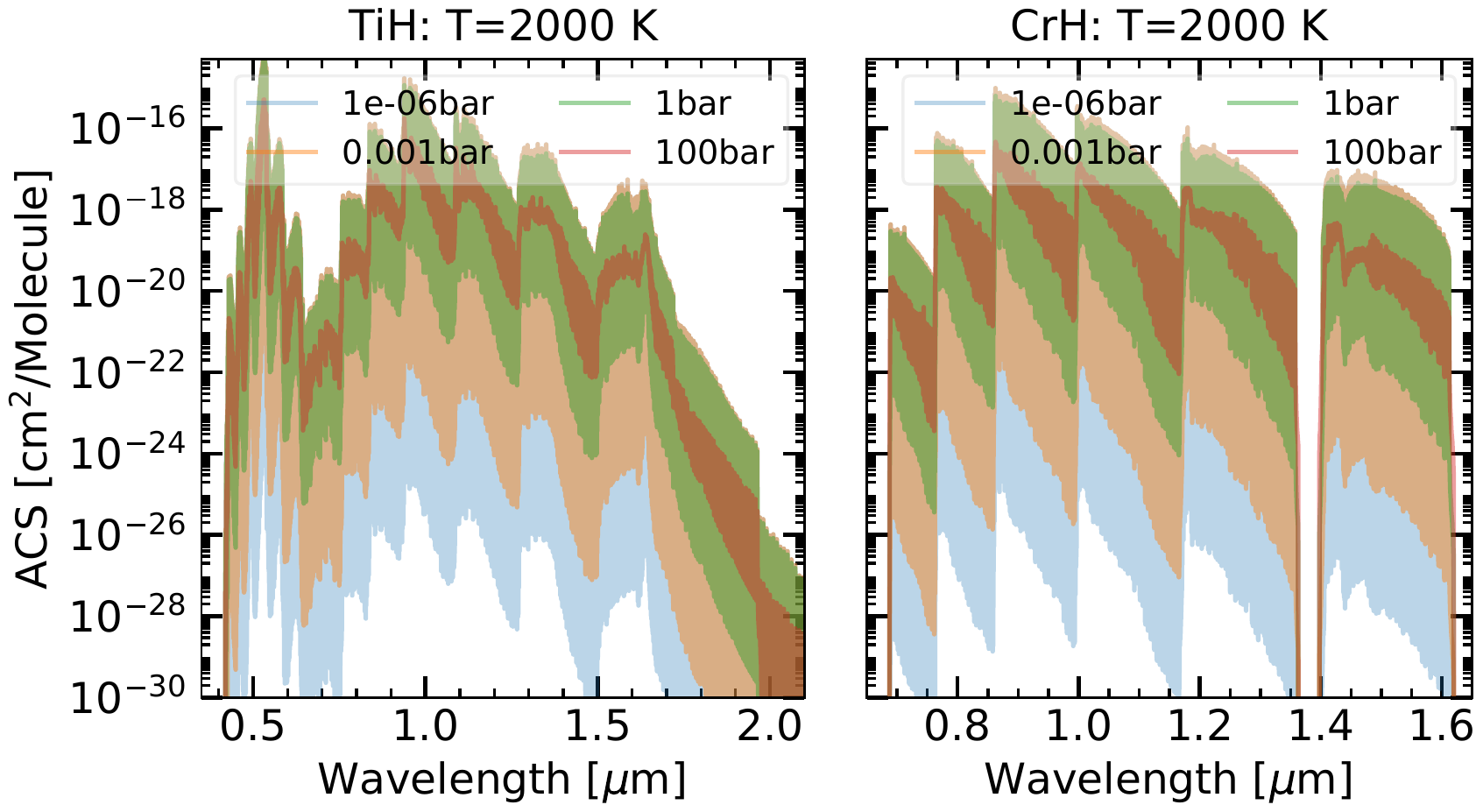} 
	  \caption{ Example of our pre-generated TiH (left) and CrH (right) absorption cross-sections data at 2000 K, and multiple pressures. The TiH line list have been generated using a combination of {\it ab  initio} data as well as spectroscopic measurements for 0.4--2.1~$\micron$ (A--X and B--X systems) by \citet{Burrows2005TiH}. The CrH line lists are also used from \citet{Burrows2002_CrH} for 0.65--1.65~$\micron$ (A--X system). These linelists are limited to these wavelength range, and future laboratory/ab-initio studies are needed to extend these spectra to higher wavelengths (e.g., for CrH at 1.38~$\micron$ gap).   
	  }
	  \label{fig:TiH-CrH-XS} 
\end{figure*}

Several spectroscopic works have been performed to record and assign the TiH spectral systems including the $^4\Gamma-X^4\Phi$ system (530~nm  \cite{Steimle1991,Launila1996}) and the vibrational $^4\Phi-X^4\Phi$ system (938~nm  \citep{Andersson2003,Linton2012}) system. \citet{Burrows2005} re-investigate the previous TiH spectroscopy data in order to derive new spectroscopic constants to produce a line list, called MoLLIST. We employed this line-list\footnote{ \href{http://exomol.com/data/molecules/TiH/48Ti-1H/MoLLIST/}{exomol.com/data/molecules/TiH/48Ti-1H/MoLLIST/}} to generate pre-computed ACS data over the parameters given in Table \ref{tab:comp}.  Cross-sections for select pressures are shown in  Fig.~\ref{fig:TiH-CrH-XS} (left).

\subsection{CrH}\label{sec:CrH}
An emission spectrum of the $A^6\Sigma^+-X^6\Sigma^+$ (0.74--1~$\micron$) system was first recorded by \citep{Ram1993}, later expanded upon by \citep{Bauschlicher2001} who fit more bands (i.e., 1--0, 1--1, and 2--0) in the region of 0.67--1.11~$\micron$. Recently, \citet{Bernath2020MOLLIST} combined them all these line-lists into the MoLLIST database. We used MoLLIST reformatted by the ExoMol group\footnote{ \href{http://exomol.com/data/molecules/CrH/52Cr-1H/MoLLIST/}{exomol.com/data/molecules/CrH/52Cr-1H/MoLLIST/}} to generate ACS data for the available spectral range (i.e., 0.7--1.6~$\micron$). Fig.~\ref{fig:TiH-CrH-XS} (right) shows our cross-sections as a function of pressure.

\subsection{CaH}\label{sec:CaH}
\begin{figure*}[!htb]
\centering
\includegraphics[scale=1.0]{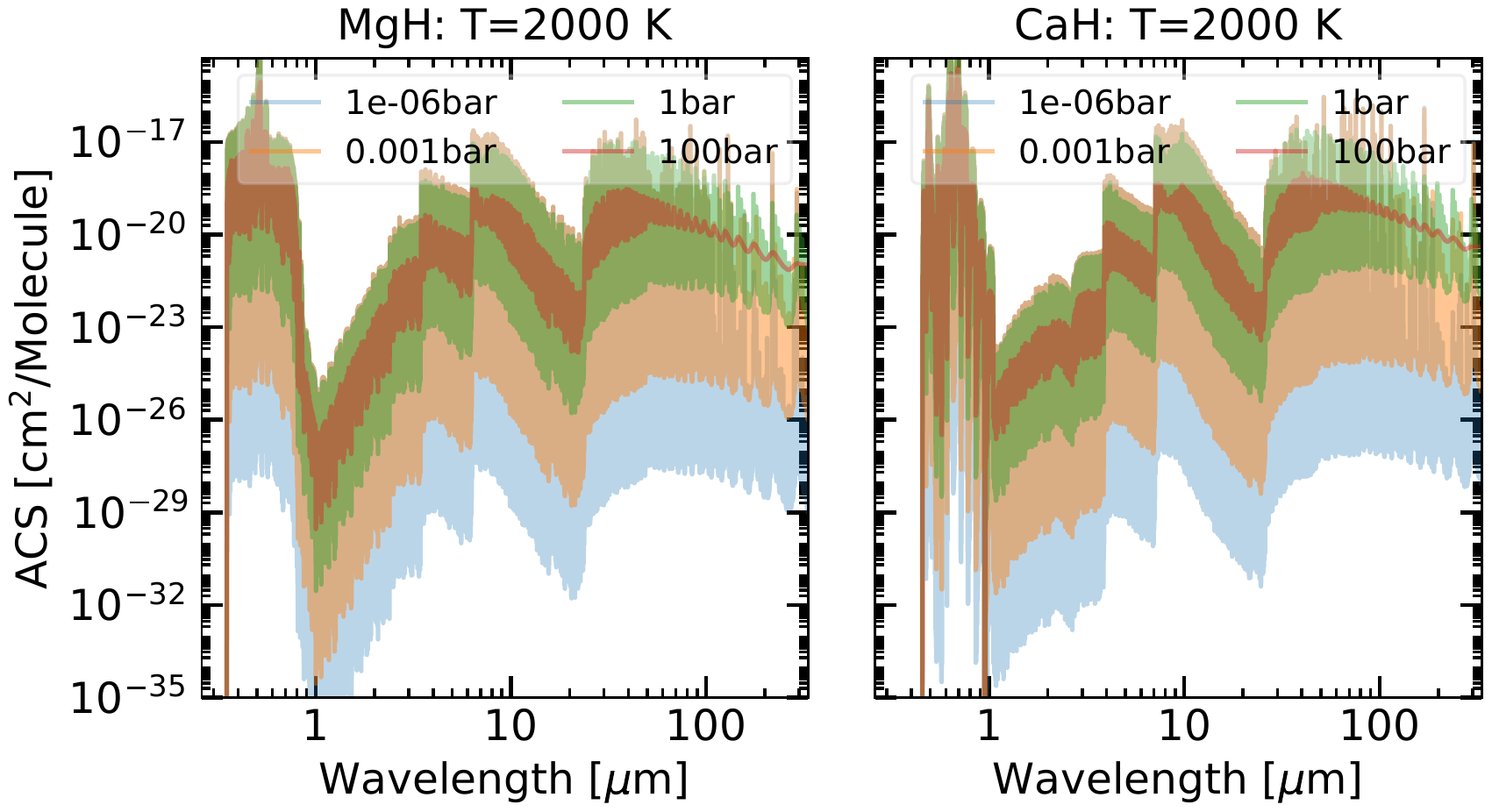} 
	  \caption{Example of our pre-generated MgH (left) and CaH (right) absorption cross-sections data at 2000 K, and multiple pressures. The generated MgH is a combination of the laboratory line list from \citet{GharibNezhad2013MgH} for 0.35--1.2$\micron$($A-X$ and $B-X$ transitions) as well as theoretical infrared line lists from \citet{Yadin2012MgH-CaH}. The CaH includes laboratory measured line lists from \citet{Alavi2017CaH, Li2012CaH} for 0.45--1.0 ($A-X$ and $B-X$), 0.7--11 (rovibrational transitions) by \citet{Shayesteh2004CaH}, and the~$\micron$ ($E-X$) transitions $\sim$0.48~$\micron$ by \citet{Li2012CaH}. CaH is also consists of theoretical line list generated by \citep{Yadin2012MgH-CaH} for rovibrational transitions. See Table~\ref{tab:Summary-linelist} for further details.}
	  \label{fig:MgH-CaH} 
\end{figure*}

The infrared and electronic transitions of the ground state $X^2\Sigma^+$, excited states ($A^2\Pi$, $B^2\Sigma^+$, and $E^2\Sigma$, 0.45--1 and 2--14~$\micron$ respectively) have been characterized using high-resolution Fourier transform spectroscopy \citep{Ram2011,Shayesteh2013}. The Einstein-A-coefficients have also been calculated for these band systems \citep{Li2012, Alavi2017, Yadin2012MgH-CaH}. Additional {\it ab initio} investigations determined the transition dipole moments of the ground state $X^2\Sigma^+$, excited states (i.e., $A^2\Pi$, $B^2\Sigma^+$, and $E^2\Sigma$) \citep{Shayesteh2017, Weck2003} covering wavelengths from 1--2 and 14--100~$\micron$.  Our cross-sections employs the laboratory-based lists synthesized by \citet{Bernath2020MOLLIST} \footnote{ \href{http://exomol.com/data/molecules/CaH/40Ca-1H/MoLLIST/}{exomol.com/data/molecules/CaH/40Ca-1H/MoLLIST}} over the 0.45--1 and 2--14~$\micron$ ~regions. For the remaining spectral range, i.e. 1--2 and 14--100~$\micron$, we used the {\it ab initio} line list from \cite{Yadin2012MgH-CaH}\footnote{ \href{http://exomol.com/data/molecules/CaH/40Ca-1H/Yadin/}{exomol.com/data/molecules/CaH/40Ca-1H/Yadin/}}.  Figure~\ref{fig:MgH-CaH} (right) summarizes our cross-sections at 2000K over a range of pressures.

\subsection{MgH}\label{sec:MgH}
The spectroscopy of the ground state and low-lying excited electronic states of MgH has been the subject of numerous investigations. \citet{Balfour1975a,Balfour1975b} and later \citet{Shayesteh2011} have measured the rovibronic transitions of the $A-X$ and $B^\prime-X$ systems in order to extract line positions, dissociation energies, rotational constants, and their potential energy surfaces. Employing the latest potential energy surface's, \citet{Yadin2012MgH-CaH} calculate all infrared transitions for the MgH ground electronic state for 1.2--100~$\micron$. In addition,  \citet{GharibNezhad2013MgH} used high-resolution spectroscopic data to produce a line list from 0.35--1.2~$\micron$.  We compute our cross-section using the \citet{Yadin2012MgH-CaH} \footnote{\href{http://exomol.com/data/molecules/MgH/24Mg-1H/Yadin/}{exomol.com/data/molecules/MgH/24Mg-1H/Yueqi/}} line list from 1.2--100~$\micron$ and the \citet{GharibNezhad2013MgH}\footnote{ \href{http://exomol.com/data/molecules/MgH/24Mg-1H/MoLLIST/}{exomol.com/data/molecules/MgH/24Mg-1H/MoLLIST}} from 0.35--1.2~$\micron$. Figure~\ref{fig:MgH-CaH}(left) illustrates a slice on our cross-section grid for 2000K over a representative range in pressures.

\subsection{SiO}\label{sec:SiO}
\begin{figure*}[!htb]
\centering
\includegraphics[scale=1.0]{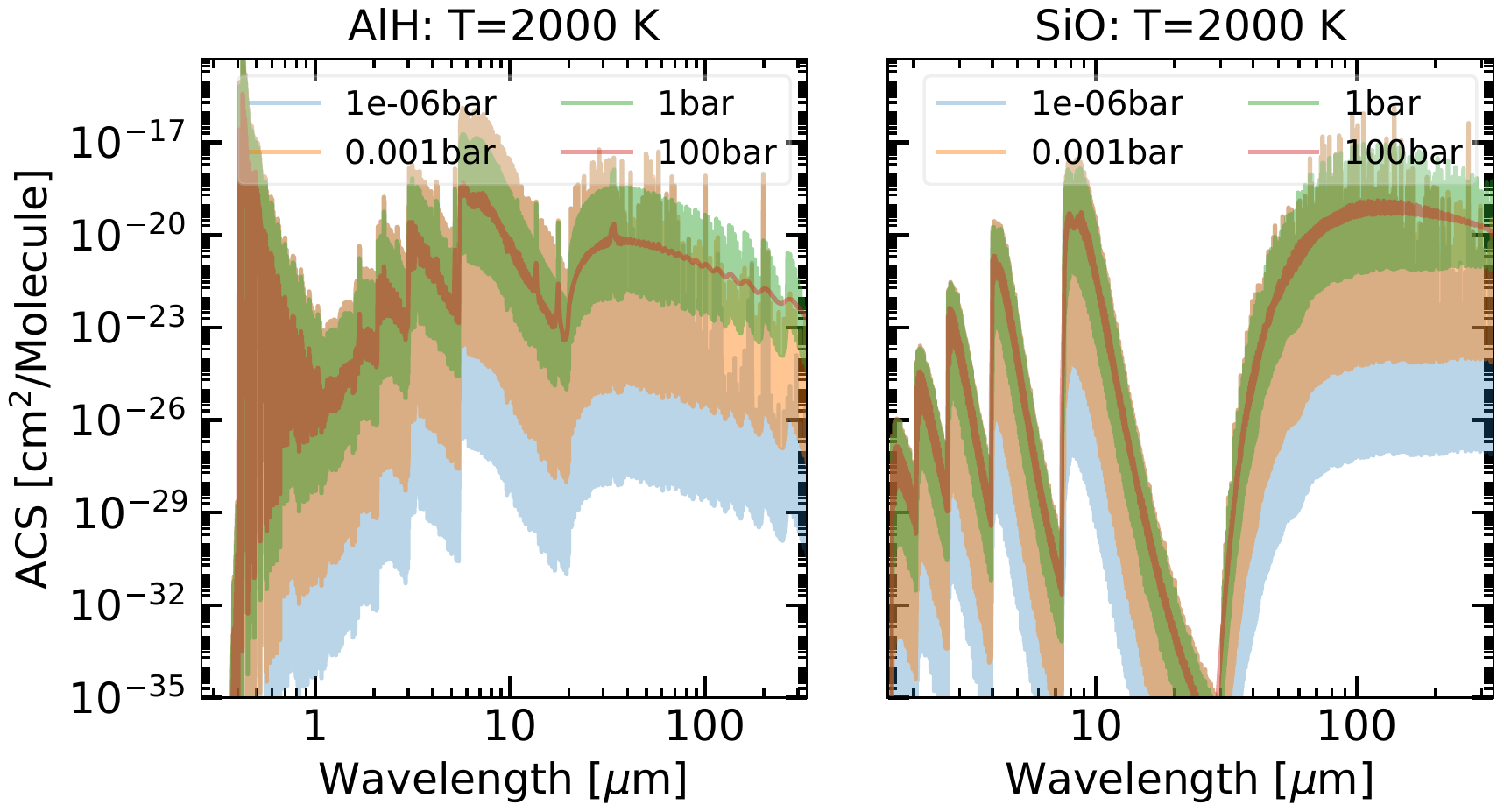} 
	  \caption{Example of our pre-generated AlH (left) and SiO (right) absorption cross-sections data at a 2000 K and multiple pressures. The AlH line list (named WYLLoT) is obtained from experimentally-calculated electronic states of $X$ and $A$ by \citet{Yurchenko2018AlH} and so consists of A--X and X--X transitions. On the other hand, the SiO line list in our study has been computed using the available recorded SiO from both laboratory and sunspot spectra as well as the computed dipole moments by \citet{Barton2013SiO}.According to \citet{Barton2013SiO}, the SiO lines above 10000 cm$^{-1}$ are very weak and so any lines between 10000--65000 cm$^{-1}$ was filtered in their study. }
	  \label{fig:SiO-AlH} 
\end{figure*}

Since the 1970s, SiO has been the subject of several spectroscopic studies over the UV-Vis to the IR using data recorded from both hot gas cells as well as the sunspots \citep{Barrow1975,Campbell1995,Sonnabend2006}. Furthermore, the SiO potential energy surfaces, Einstein A coefficients, and other spectroscopic constants have been theoretically studied using {\it ab inito} method \citep{Langhoff1993,Langhoff1979}.  Later, as a part of the ExoMol initiative, \citet{Barton2013SiO} improved the potential energy surface of the ground state and dipole moment function. In our calculation, we used the \citet{Barton2013SiO} line list\footnote{ \href{http://exomol.com/data/molecules/SiO/28Si-16O/EBJT/}{exomol.com/data/molecules/SiO/28Si-16O/EBJT/}} to calculate cross-sections over 1.67--100~$\micron$ for the temperature and pressure range’s reported in Table~\ref{tab:comp}. Figure~\ref{fig:SiO-AlH}(right) shows a slice on our ACS grid for 2000~K over a representative range in pressures.

\subsection{AlH}\label{sec:AlH}
Different rovibronic lines of AlH have been measured in the lab using high-resolution spectroscopy, including the $A^1\Pi-X^1\Sigma^+$ band at 400--550~nm \citep{Szajna2009,Ram1996AlH}, the $C^1\Sigma^+-X^1\Sigma^+$ band at 220--240~nm \citep{Szajna2010}. In addition, the AlH dissociation energy and the low-lying electronic states have been characterized using {\it ab initio} approach’s \citet{Matos1987}. Recently, \citet{Yurchenko2018AlH}  employed these spectroscopic constants of the ground state and the first low-lying excited state, and experimental potential energy curves to calculated all rovibrational transitions between the energy levels. We employed their line list, named  $WYLLoT$\footnote{\href{http://exomol.com/data/molecules/AlH/27Al-1H/AlHambra/}{exomol.com/data/molecules/AlH/27Al-1H/AlHambra/}}$^,$\footnote{Although in the paper by \citet{Yurchenko2018AlH}, it was named as WYLLoT line list, the ExoMol website provided the AlH line list files with ``AlHambra’’ name.}, to generate our ACSs from 0.4--100~$\micron$ for the temperature and pressure ranges in  Table~\ref{tab:comp}. Figure~\ref{fig:SiO-AlH} (right) represents a subset of our ACSs as a function of pressure.

\subsection{Comparison to Other Absorption Cross-Section Databases} \label{sec:compare-to-other-opacities}
There are other cross-section databases, { some publicly available such as \citep{VILLANUEVA2018OpacityGenerator, Chubb2020}, and some are not 
\citep{Sharp2007, Freedman2008, Freedman2014, Goyal2018ATMO}}. We build on these past works by implementing the latest line lists, primarily drawn from the ExoMol, and provide an up-front description of our sources of pressure broadening and how we approach the problem when no data is available (see \S\ref{sec:XS-background}). Table \ref{tab:Summary-compareXS} summarizes the known differences between our database and past absorption cross-section database's. In summary, our database resolution is sufficiently large to be applicable for interpretation of $HST$ and $JWST$ observed data as well as high-resolution cross-correlations'. In addition, we used a constant number for the wing cut-off to avoid overestimation of the opacity continuum. The temperature, pressure, and wavelength ranges are also wider and have more grids (1460 T-P points) than other available pre-generated databases in order to fully model the (Ultra)Hot-Jupiters and M-L dwarf atmospheres.

\section{Impact of Opacities on highly-irradiated exoplanet atmospheres}\label{sec:Impact-on-RTmodeling}

We assess the impact of the newly generated ACSs on the atmospheres and spectra of  M dwarfs and hot Jupiters--specifically, the influence of metal hydrides and oxides from {\tt EXOPLINES}  vs. Freedman2014 \citep{Freedman2014}{ (see the list of the Freedman2014’s line lists at Table 2 by \citet{Lupu2014EarthLike})}. First, we simulate a representative Ultra-Hot Jupiter atmosphere (T$_{irr}$= $\sim$2000 K,  $\log g$=3, T$_{star}$=6500 K, solar composition) with a 1D-self-consistent radiative-convective thermochemical equilibrium model, {\tt Sc-CHIMERA}\footnote{The core radiative transfer, {\tt CHIMERA}, found at \hyperlink{https://github.com/mrline/CHIMERA}{https://github.com/mrline/CHIMERA}}, previously described in \cite{Piskorz2018hotJupiters, Arcangeli2018, GharibNezhad2019}. We compute the planetary emission and transmission spectrum from 0.3--5.0~$\mu$m. Major opacity sources included in these models are  \ce{H2O}, TiO, VO, FeH, CO, CO$_2$, CH$_4$, NH$_3$, H$_2$S, HCN, C$_2$H$_2$, PH$_3$, CaH, MgH, CrH, SiO, AlH, Na, K, \ce{H2}--\ce{H2}, \ce{H2}--He, Fe, Mg, Ca, H- bound/free-free continuum, and H$_2$/He Rayleigh scattering within the correlated-K resort-rebin framework \citep{Amundsen2016}.

\begin{figure*}[!htb]
\centering
\includegraphics[width=\textwidth]{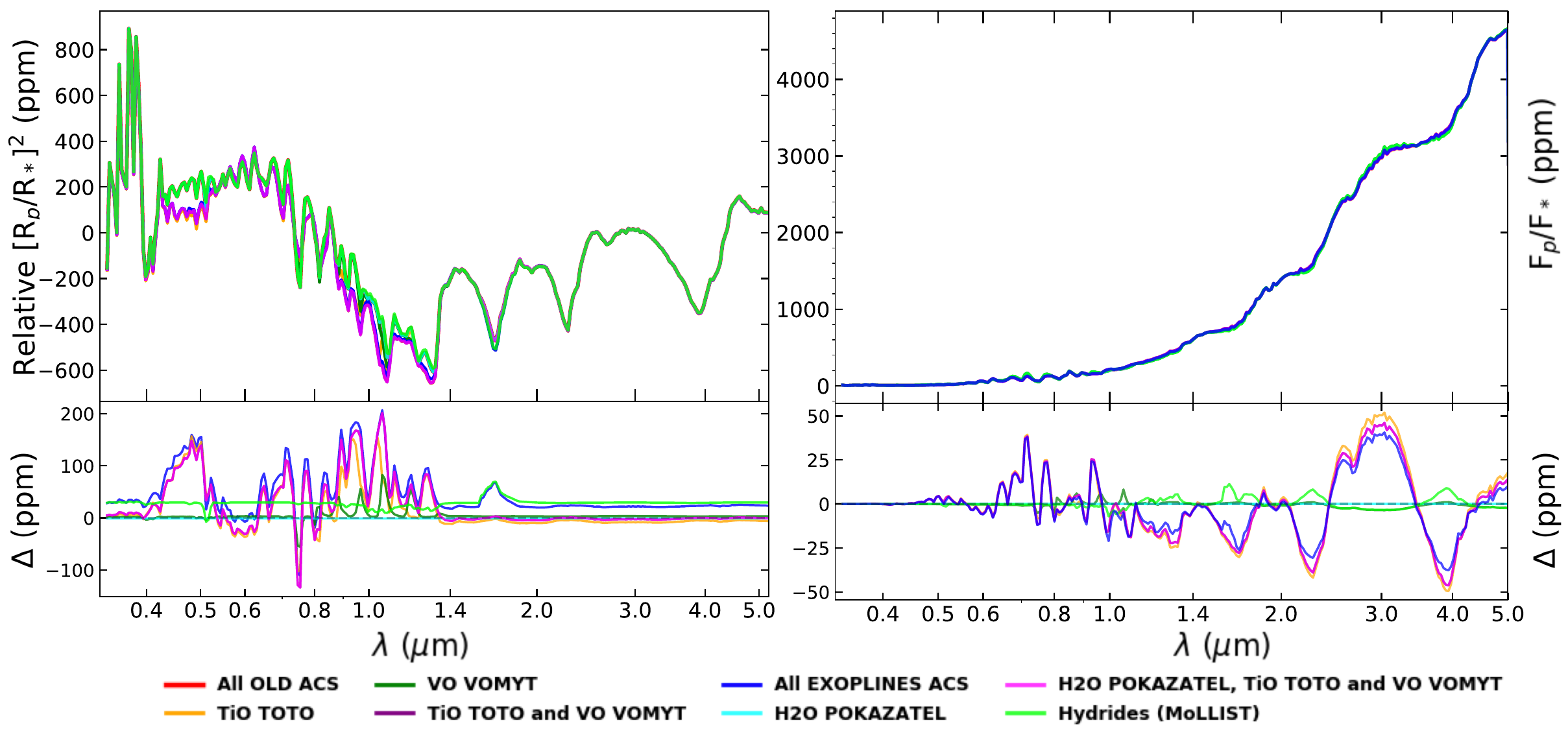} 
	  \caption{Impact of pre-generated opacities in the transmission and emission spectrum of a representative Ultra-Hot Jupiter. Transmission (Left) and Emission (Right) spectra calculated using opacities from all old ACSs (red) compared to all {\tt EXOPLINES}  ACSs (blue). Choice of linelists for TiO (TOTO vs Schwenke, in orange), VO (Plez vs VOMYT, in green), \ce{H2O} (Schwenke vs POKAZATEL, in cyan), and metal hydrides (in lime, see Table \ref{tab:Summary-compareXS} references) along with combination differences of metal oxides (TiO and VO, in purple) as well as \ce{H2O}(magenta) showing notable residuals relative to all old ACS data (red) of up to 300 ppm, primarily in the optical bandpass, between 0.3-1.0 $\mu$m at $JWST$ fiducial resolution at R=100 for transmission (Bottom Left) and up to 80 ppm in the infrared between 1.0-5.0 $\mu$m for thermal emission (Bottom Right), despite no significant shape differences amongst the spectra. In both cases, TiO produces the largest residual differences. Therefore, significant bias may arise with the choice of linelist in model atmospheres, especially when characterizing ultra hot Jupiter atmospheres with $JWST$.}
	  \label{fig:WASP121b-emissionTransmission} 
\end{figure*}

We illustrate the effects of TiO, VO, \ce{H2O} individually, and as combined with other metal hydrides (FeH, CaH, MgH, CrH, SiO, and AlH, with sources listed in Table \ref{tab:Summary-linelist}). Figure \ref{fig:WASP121b-emissionTransmission} shows both transmission (left) and emission (right) spectra for an Ultra Hot Jupiter, varying only the choice of line list. Focusing first on the transmission spectrum, we report significant differences in the spectral shape below 1.0$\mu$m; however, in the near-infrared, the residual differences are well below 50 ppm\footnote{$HST$ WFC3 (1.1--1.4~\micron)~precision of 15--50 ppm are routinely achieved \citep{Line2016, Kreidberg2014}. Comparable or better precision are anticipated with $JWST$ \citep{Bean2018JWST}    }. Between 0.3--1.0~$\mu$m, we find that major differences are due to the choice of TiO line list (TOTO vs. Schwenke, \cite{Freedman2014}), leading up to 320 ppm in residuals. Differences arising purely from the choice of metal hydride line lists (FeH, MgH, and CaH from MoLLIST), are around 75 ppm for the transmission spectrum, notably over wavelengths covered by the HST/WFC3 bandpass (1.4--2.0~$\mu$m). We also investigated the influence of \ce{H2O} line list choice on the transmission spectra, where we find residual differences fall under 10 ppm when comparing Schwenke \citep{Freedman2014} to $POKAZATEL$. 

For the thermal emission spectrum; similarly, TiO line list choice leads to significant differences in the spectral shape, especially from 1.0--5.0~$\mu$m, with up to 80~ppm differences. Metal hydrides and \ce{H2O} lead to residual differences around 10 ppm, also consistent with the transmission spectrum (see Figure \ref{fig:WASP121b-emissionTransmission}). Moreover, the total differences between all old ACS data (e.g., TiO-Schwenke, VO-Plez, MgH-Weck, CaH-Weck, FeH-old) against the {\tt EXOPLINES}  ACS data (TiO-TOTO, VO-VOMYT, MgH-MoLLIST, CaH-MoLLIST, FeH-MoLLIST), are dominated by TiO and metal hydride line list differences, altogether leading up to the same level of residuals. 

\begin{figure}[!htb]
\centering
\includegraphics[scale=1.0]{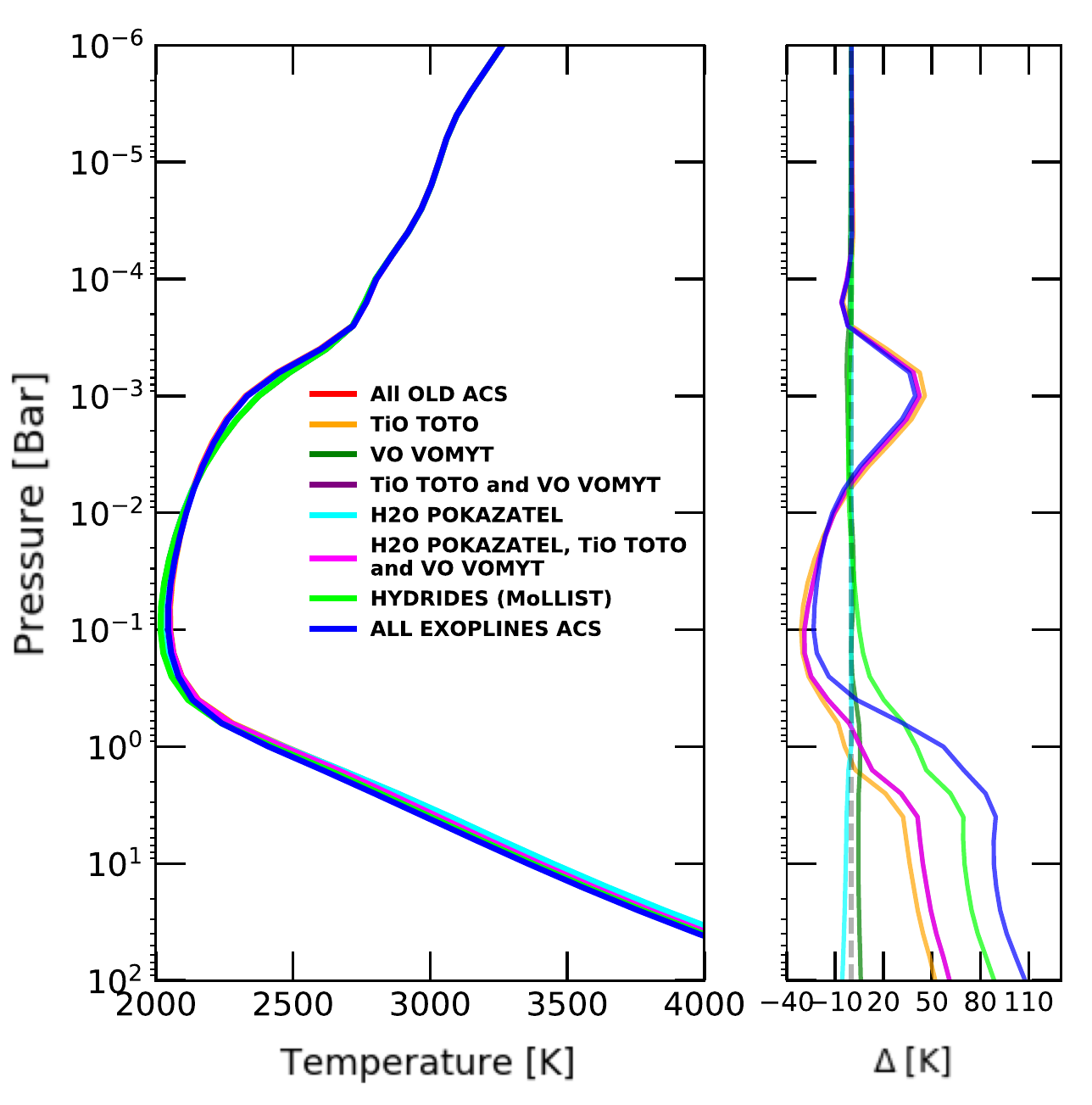} 
	  \caption{The impact of pre-generated opacities in the atmospheric structure of a representative Ultra-Hot Jupiter.(Left) Thermal structure calculated using pre-generated opacities from all old ACS (red) (\cite{Freedman2008}, See Table \ref{tab:Summary-compareXS}), compared to {\tt EXOPLINES}  (blue). (Right) Independent effects of TiO (TOTO), VO (VOMYT), \ce{H2O} (POKAZATEL), metal hydrides (FeH/MgH/CaH/CrH-MoLLIST, and AlH-WYLLoT) (orange, green, cyan and lime respectively) as well as combination differences of all {\tt EXOPLINES}  ACS (blue) show variation in the thermal profile of up to 130 K for the same planet. The largest variations in the thermal profile are purely due to ACS differences of metal hydrides and oxides that are dominant in highly irradiated planet atmospheres.}
	  \label{fig:WASP121b-TP} 
\end{figure}

We also illustrated the impact of these new ACSs data on the Temperature-Pressure (TP-) profile of this candidate WASP-121b atmosphere in Fig.~\ref{fig:WASP121b-TP}. This figure shows that the new TiO-TOTO ACS data change the TP profile by $\sim$60~K between 1mbar and 1 bar. The impact of all new molecular ACSs (TiO, VO, etc.) is $\sim$130~K between 1~mbar to 1~bar. These differences in the TP-profile lead to change in the atmospheric chemical composition, emission, and transmission synthetic spectra.  

 More recently, \citet{Piette2020TiO-UHJ} showed that incorporating the different TiO opacities (i.e., TOTO, Plez, and Schwenke) biases the synthetic emission and transmission spectra of ultra-hot Jupiters (R$\sim$10$^{4}$) up to $\sim$1000~ppm in the visible and 100~ppm in the near-IR spectral region. In another instance for high-resolution ground-based spectroscopy, \citet{Herman2020TiO-Wasp33b} used TiO-Plez ACS to identify the potential presence of TiO optical lines in WASP-33b. Implications of these works strongly vouch for the critical vetting of line lists used to compute appropriate ACS, especially concerning oxygen-bearing molecules such as \ce{H2O} and TiO to robustly determine bulk chemical properties (such as C/O ratio) of sub-stellar atmospheres \citep{Molaverdikhani2019irradiated}, as well as improving the atmospheric circulation models of hot Jupiters \citep{Kataria2016}.

\section{Impact of Opacities on M-dwarf atmospheres}\label{sec:Impact-on-RTmodeling-Mdwarf}
\begin{figure*}[!htb]
\centering
\includegraphics[width=\textwidth]{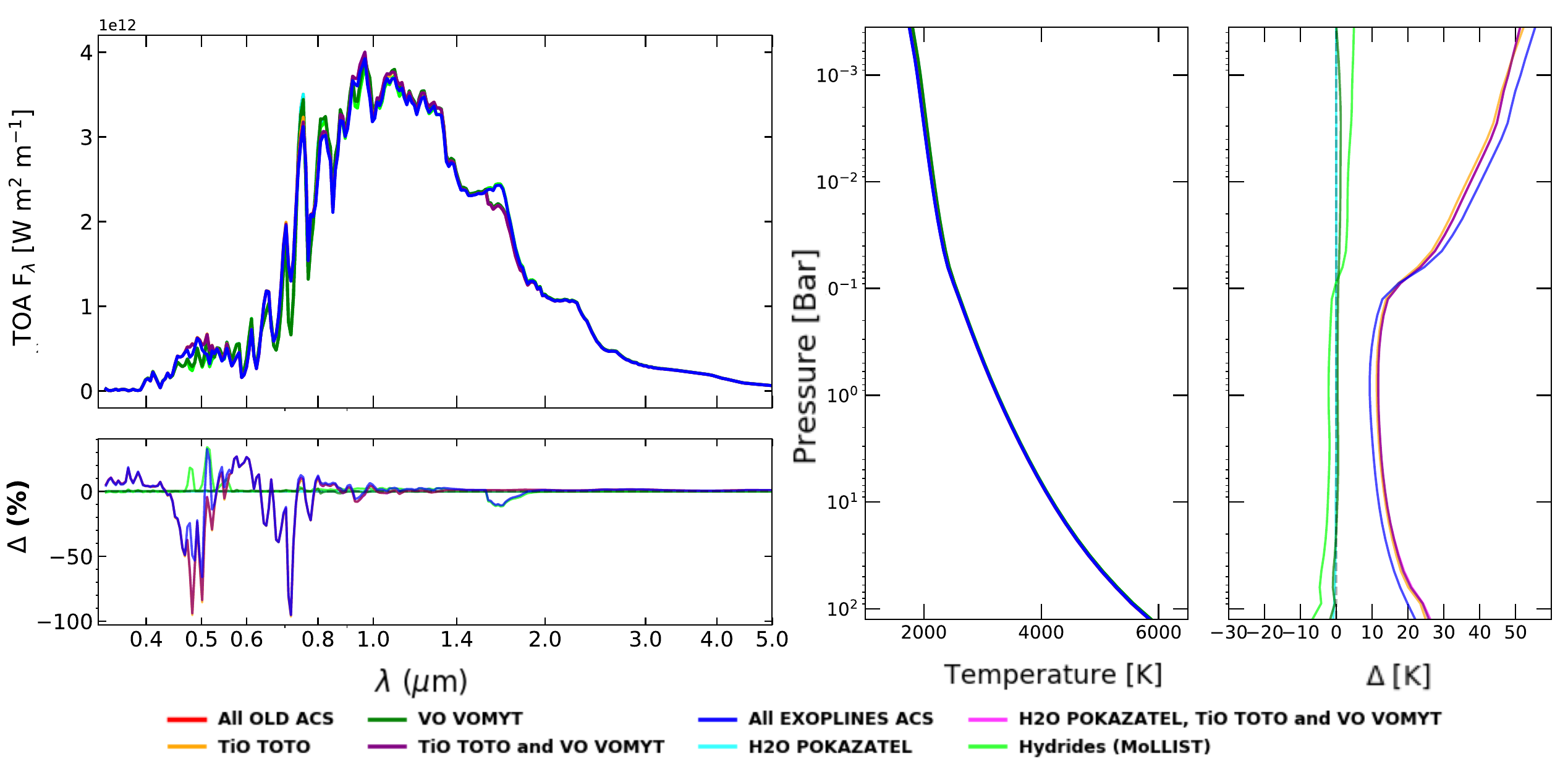} 
	  \caption{Impact of pre-generated opacities in characterizing a fiducial M-dwarf atmosphere, with T$_\mathrm{eff}$ of 3000 K, $\log\mathrm{(g)}$ of 5 cm.s$^{-2}$, solar M/H, and C/O of 0.5. (Left) Model spectrum of the fiducial M-dwarf (showing top-of-atmosphere (TOA) flux) varying opacities calculated from all old ACS (red) compared to all {\tt EXOPLINES}  ACS (blue). Spectrum (R=100, covering 0.3-5$\mu$m) calculated with TiO TOTO linelist (orange) attributes to largest residual differences in the spectral shape, with up to 125$\%$ in the optical bandpass below 1$\mu$m, followed by 30$\%$ in residual differences caused purely due to the effect of metal hydride opacities (MoLLIST, lime) below 1$\mu$m, and up to 12$\%$ between 1.4-2.0$\mu$m. \ce{H2O} (POKAZTEL, cyan) and VO (VOMYT, green) alone do not induce noteworthy differences within the spectra. (Right) Atmospheric structure of fiducial M-dwarf, with respective residuals in the thermal profile. Consistent with the spectral differences, TiO linelist contributes to significant deviations within the T-P profile with up to 50 K above 0.1 bar in pressure, and up to 20 K down to 100 bars. The effect of both metal oxides (TiO and VO, purple), combined with \ce{H2O} (magenta) and along with all new opacities (EXOPLINES, blue) cause deviations in atmospheric structure of up to a maximum of 60 K.}
	  \label{fig:mdwarf_opacitydiff} 
\end{figure*}

M-dwarf atmospheric spectra are also subject to differences arising from various line lists. To quantify these differences, using the same self-consistent modeling framework described above, we simulate a dust-free $T_{\mathrm{eff}}$ = 3000~K, $\log g$=5, solar composition atmosphere. In an upcoming paper by \citet{iyermdwarf1}, we will discuss our M-dwarf model in great detail. Consistent with the hot Jupiter case, TiO line-list differences have the largest influence on the spectral shape and the temperature profile. We find that the TiO-$TOTO$ ({\tt EXOPLINES}) line-list alone, when compared to the $Schwenke/Allard$ (Figure \ref{fig:mdwarf_opacitydiff}, left) produces spectral differences of up to 125$\%$, primarily between 0.3--1.0~$\mu$m. Alternatively, the effect of VO opacity ($Plez$ \citep{Freedman2014} vs $VOMYT$ ({\tt EXOPLINES})) and \ce{H2O} (Schwenke \citep{Freedman2014} vs $POKAZATEL$ ({\tt EXOPLINES})) do not contribute to any notable differences at lower resolutions (R=100), however, we see up to a 30$\%$ difference due to the effect of metal hydrides below 1$\mu$m, and up to 12$\%$ over the near-IR (1.4--2.0~$\mu$m) bandpasses. TiO attributes to the largest variation in the temperature profile as well; (see Figure \ref{fig:mdwarf_opacitydiff}, right), with up to 50 K in differences between 1 mbar to 0.1 bar, and up to 20 K difference deeper than 100 bars, because of a subtle shift in the adiabat. Furthermore, the absorption cross-sections from the hydrides (calculated from MoLLIST) lead to thermal profile differences of up to 10 and 20 K respectively, from 1 mbar to 0.1 bar and down to 100 bars.
Undoubtedly, this level of bias---in both hot Jupiters and M-dwarfs---arising purely from generational differences of line lists would lead to improper atmospheric inferences. \citet{McKemmish2019TiO-TOTO} also have recently investigated the impact of \citep{Plez1998}, \citep{Schwenke1998TiO}, and TOTO TiO line lists on M-dwarf flux, which showed large differences in their synthetic flux.

\section{Current Challenges with Opacities} \label{sec:opacity-challenge}
\begin{figure}[!htb]
\centering
\includegraphics[scale=0.55]{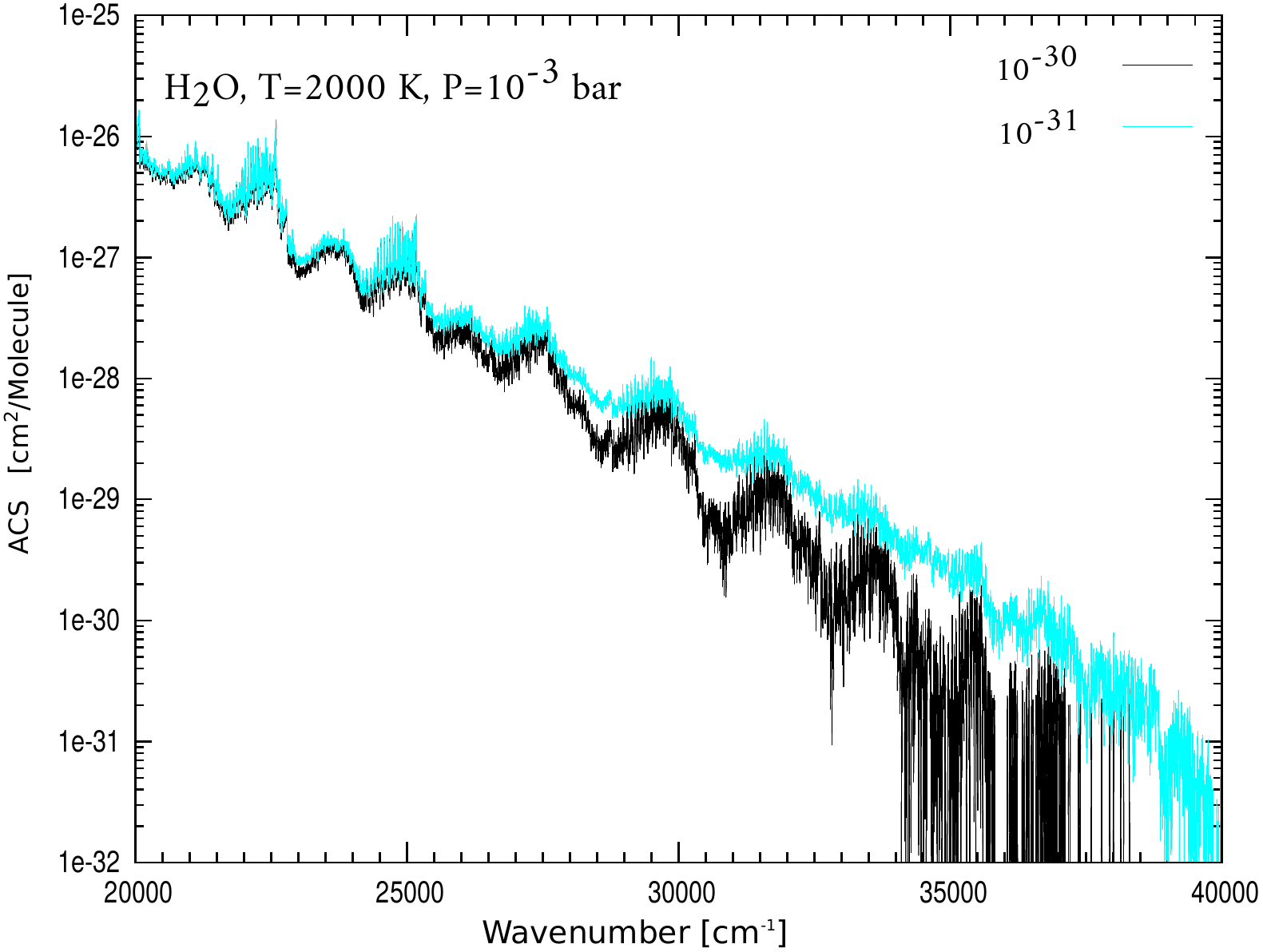} 
	  \caption{The impact of intensity cut-off on generating the \ce{H2O} $POKAZATEL$ ACS data. Following Maxwell-Boltzmann distribution, the population of energy levels with high $J$ quantum number increases with temperature. As a result, weak lines become stronger at high temperatures, and so insufficient intensity cut-off value results in accurate ACS. For instance, the choice of 10$^{-30}$ (black) for intensity cut-off eliminate more lines in generating ACS data comparing to 10$^{-31}$ cm$^{-1}$/(molecule cm$^{-2}$) (cyan). The reported cut-off values refer to  296 K.}
	  \label{fig:intensity-cutoff} 
\end{figure}

Several spectroscopic parameters are critical for computing absorption cross sections and their subsequent effect in the modeling of observations \citep[e.g.,][] {Fortney2019whitepaper}. Some challenges arise simply due to lack of laboratory data, and some are because of the computational shortcuts used to improve the speed of these often lengthy calculations. Therefore, it is crucial to consider these all differences between the Voigt profiles, broadening treatment, wing cut-off and other details in order to evaluate and cross-check different opacity databases generated by either CPU-based computing codes such as ExoCross and HAPI\footnote{\href{https://github.com/hitranonline/hapi}{https://github.com/hitranonline/hapi}}, or GPU-based codes such as   Radis\footnote{\href{https://radis.readthedocs.io/en/latest/}{https://radis.readthedocs.io/en/latest/}} and Helios-K\footnote{\href{https://helios-k.readthedocs.io/en/latest/}{https://helios-k.readthedocs.io/en/latest/}}\citep{yurchenko2018exocross, Kochanov2016HAPI, Bekerom2021Radis, Grimm2021}. We briefly discuss the current pressing challenges in producing accurate cross-sections:

\begin{itemize}[label=\tiny$\blacksquare$,leftmargin=0.3cm]
\item { The Lack of Pressure Broadening Data---As discussed in \S\ref{sec:XS-background}, the pressure-broadening coefficients, $\gamma_\mathrm{L}$ and $n_T$, are critical to the proper computation of the Lorentz line widths and wings, which dictates the ``off line" continuum. For an absorber, these coefficients depend on the broadening gas, quantum $J$ number, molecular symmetry\footnote{Spectroscopic parameters such as pressure-broadening and molecular angular momentums (simply, quantum numbers) depend on the molecular geometry and the symmetry between different energy states. For example, \ce{CH4} is a tetrahedral molecule and has $A_1, A_1, F_1, F_2$ and $E$ symmetries, which can be found from character tables. For further details, check Table 3 by \citet{rothman2005hitran} and the discussion in chapter 3 by \citet{Bernath_book}. Several examples of broadening data can also be found in chapters 4 and 5 by \citet{Buldyreva2011book}. }, temperature, and also the specific interactions with the broadening gas (e.g., the complexity in the line profile of Na and K atoms studied by \citet{Allard2019Na-H2,Allard2016K-H2}). The current state of broadening data for many molecules important in exoplanet/substellar atmospheres is concerning. In most cases for many absorbers $\gamma_\mathrm{L}$ measurements are limited to below 500 K (e.g., see review by \citet{Hartmann2018review}). Most data are only applicable to ``Earth-like" conditions through “air-broadening” or reducing solar-system giant-like atmospheres through H$_2$/He broadening. Regarding the metal hydrides and oxides in question, as far as our best knowledge, there is no laboratory/theoretical broadening data available.. }

\item { Compositional dependency of absorption cross sections: One of the key findings of the Kepler Mission is that a majority of exoplanets fall within this ``warm
mini-Neptune’’ regime (R$_p$ $\sim$2--4R$_{Earth}$ T$<$1000 K) \citep{Batalha_Nat2014}. The bulk composition of these exoplanets is not hydrogen-dominant, and therefore, incorporating the pre-generated \ce{H2}/He-broadened opacities to model their atmospheres may cause errors and biases in interpreting their observational emission and transmission spectra. For example, \citep{GharibNezhad2019} have shown that using the \ce{H2}/He-broadened of water opacity for modeling a candidate super-Earth steam atmosphere could lead up to 80 ppm (part-per-million) difference in transmission and 250 ppm in the emission spectra, which are detectable by both $HST$ and the future $JWST$ telescopes. As a result, there is uncertainty and inaccuracy in atmospheric modeling observables such as composition, which could lead to a misinterpretation of observational data.
}
\item Line Profile---The Voigt profile is the most common line shape applied in opacity computations. This profile is based on the ideal gas assumption where no interactions between broadening (e.g., the bath gas/background) molecules exist. However, these interactions do occur, and hence, this effect should be taken into account with more advanced line profiles such as the speed-dependent Voigt profile \citep{Pickett1980speedDependent, Buldyreva2011book}. These broadener gas interactions have been explored at both low and high pressures \citep{Ngo2012H2O_nonVoigt-profile, GRIGORIEV1999}. Table 1 by \citet{NGO2013Non-Voigt} illuminates some of these differences.

\item Voigt Profile Algorithm---The specific algorithm and inherent approximations used to compute the Voigt profile can result in non-negligible uncertainties. The Voigt algorithm by \citet{Humlicek1979} is the most widely used due to the balance between accuracy and computational efficiency. However, it has been shown \citep{zaghloul2015Voigt, Schreier2018Voigtchallenges} that this algorithm is not always accurate over the pressure, temperature, and wavelength regions. Other algorithms exist, but need to be continually assessed for accuracy in different situations \citep{Schreier2018VoigtError}.

\item Line Wing Cut-Off---This dictates the frequency extent to which the Lorentz line wings are computed from the line core--typically employed to reduce the computational burden. Choosing too short of a wing cut-off, particularly for high pressures ($>$10 bar), can result in large errors in the opacity continuum depending upon the number of transitions. \citet{Hartmann2002wing} proposed an empirical correction factor for the wing cut-off to fully model the 3.3~$\micron$ band of \ce{CH4}, however, this study is only limited to room temperature and the broadener is \ce{H2} molecule. Further investigations are required to understand the accurate wing cut-off for different line profiles.

\item Intensity Cut-Off---This effectively limits the number of lines over which to compute the line strengths and profiles, and subsequent cross sections. Line intensities are typically given as the line strength at a particular reference temperature, usually 296K. Choosing a cut will remove the weaker lines that can present as stronger lines at higher temperatures, resulting in a critical loss in overall opacity. The cutoff choice is particularly influential with line lists that contain billions of transitions, which can result in the exclusion of a substantial portion of the weaker transitions, resulting in an altered ``pseudo-continuum" shape. Figure~\ref{fig:intensity-cutoff} illustrates the impact of line-strength cut-off on the $POKAZATEL$ \ce{H2O} cross-sections at 2000 K, 10$^{-3}$ bar over the short wavelengths (0.25 - 0.5~$\micron$ ).

\item Pressure-shift--- As a result of collision by the broadening gas, the line-center blue/red shifts by $\sim$0.001-0.005 cm$^{-1}$/bar, depending on the broadener, temperature, and quantum number $J$. At high pressures, lines extensively overlap due to their large line-width, but the line center will shift as well. To the best of our knowledge, this concept has not been applied in generating ACSs for the majority of absorbers (including the current study).

\item Partition function--- The strength of each individual line for different temperatures is scaled directly by the partition function, usually treated as a separate data source in opacity calculations. Partition function value for a given temperature is calculated from all the energy levels that have been obtained from spectroscopy measurements or through $ab initio$ calculations (e.g., see equations by \citet{Barklem2016PF} and \citet{yurchenko2018exocross}). Hence, incompleteness or inaccuracy in the measured/computed energy levels results in errors in the partition function, and consequently in the computed line strengths, and resulting cross-sections. For example, the water partition function and its associated uncertainty have been studied by \citet{Vidler2000-waterPF}. In addition, \citet{SousaSilva2014} have assessed the potential errors and challenges associated with calculating the \ce{PH3} and \ce{NH3} partition function values. 
\end{itemize}

\section{Summary and Conclusions}\label{sec:conclusion}
The major objective of this investigation is to provide a comprehensive pre-computed absorption cross-section database relevant to the spectral modeling of atmospheres of irradiated hot Jupiters and substellar objects. To this end, we present a public database of pre-computed molecular absorption cross-sections for all the isotopologues of \ce{MgH}, \ce{AlH}, \ce{CaH}, \ce{TiH},  \ce{CrH}, \ce{FeH}, \ce{SiO}, \ce{TiO}, \ce{VO}, and \ce{H2O}. The pressure and temperature range of the grid span 10$^{-6}$--3000 bar and 75--4000 K, respectively (except \ce{H2O}  which is 500--4000 K) and their spectral coverage vary depending on the availability of the line list data, typically covering 0.25--330 $\mu$m where possible. 

We applied detailed care when determining the pressure broadening coefficients. Where available, we started from the published values estimated for the H$_2$/He broadening, or otherwise used a $J$-dependent collision theory. We described the history of the line lists of various species, and when possible compared the influence of available line lists on the absorption-cross sections.  Finally, we determined the impact of line list choice (hence cross-sections) on the temperature structure and emergent spectra of a hot Jupiter and M-dwarf using a 1D self-consistent radiative-convective model. From the latter experiment we found that:

\begin{itemize}[label=\tiny$\blacksquare$,leftmargin=0.3cm]
    \item The TiO-$TOTO$ line list compared to the TiO-Schwenke line-list (and subsequent cross-sections) had the most significant influence on the thermal structures of the hot Jupiter ($\sim$60K differences) and a generic 3000 K M-dwarf (50~K differences). The resultant hot Jupiter spectra (R=100) differed between the two line-list scenarios of up to $\sim$80 ppm in emission/secondary eclipse and $\sim$320 ppm (near 0.3--1~$\micron$) in transmission. The M-dwarf emergent spectra differed up to 125$\%$ over the wavelengths most influenced by TiO (0.3--1.0~$\micron$).

    \item Considering the large differences in the number of transitions in the various \ce{H2O} line lists (see  Table~\ref{tab:H2O-linelists} and Fig.~\ref{fig:H2O-compareXS-DifferentLinelist}), the resulting differences in the emergent spectra and temperature structure were are negligible at these resolutions (R=100). However, at high resolution (R$>$50,000) these differences in line positions and intensities are more apparent and have been previously shown to influence molecular abundance determinations of transiting hot Jupiter’s using cross-correlation spectroscopy \citet{Brogi2019, Gandhi2020XS-HighRes}.

    \item The metal hydride (FeH, MgH, and CaH) cross-sections derived from the MoLLIST line-lists compared to past line lists (MgH-Weck, CaH-Weck) also result in notable differences in atmospheric observables, with up to an 80~K difference in the hot Jupiter temperature structure and 10~K in the M-dwarf. Differences in the hot Jupiter emission/transmission spectra approach $\sim$50 ppm (around 1.4--2~$\micron$). In the M-dwarf spectra, up to 45$\%$ change in flux occurs near 1.0~$\micron$ and 12$\%$ between 1.4--2.0~$\micron$.

\end{itemize}

With the upcoming higher fidelity characterization of extra-solar planets and their hosts (e.g., M-dwarfs), accurate absorption cross section ingredients (line lists, broadening) are becoming increasingly critical to the success of properly estimating their atmospheric properties.  We encourage the community to continue to invest resources in line list validation and vetting, {\it ab initio} and laboratory measurements of key broadening coefficients, and production and curation of vetted absorption cross-section databases.

\section{Supplementary data}

The pre-generated absorption cross-sections (ACS) data from this study are uploaded in ZENODO \href{https://zenodo.org/record/4458189}{https://zenodo.org/record/4458189}. Our full isotopologue-weighted binary database is approximately 0.5~Tb which includes 10 molecular sets of 1460 ACS temperature-pressure data with the size of $\sim$44GB per molecule. The un-weighted individual isotopologues ACS binary data are consist of 22 sets of molecular ACS data with 1460 temperature-pressure points for each and the total size of $\sim$1~TB.     
The database will available in multiple formats to suit various needs. First, we will post a 5 Gb resampled version of the database on Zenodo, due to their file size limitations. The Zenodo database will be in the same format as the \citep{Freedman2014} database currently used by \texttt{PICASO} \citep{Batalha2019Picaso,batalha2020picaso}\footnote{\href{https://zenodo.org/record/3759675\#.Xvd0hZNKjGI}{Zenodo Opacities \citep{Freedman2014}}}. Second, the full unsampled database (MySQL format) will be available via an FTP-like portal. Last, we will also launch a web portal so that users can visualize and plot subsets of the data without needing to download it all. 

\section{Acknowledgements}

 EGN would like to thank Drs. J. Fortney, R. Lupu, I. Gordon, and R. Hargreaves for their useful comments and discussions during this study.  EGN and ARI acknowledge Research Computing at Arizona State University for providing HPC resources that have contributed to the research results reported within this paper. EGN,  MRL, and RF acknowledge support from HST-AR-15796. MRL and ARI acknowledge support from NASA grant NNX17AB56G. NEB and RF acknowledge support from 19-NUP2019-0011. We also acknowledge the ExoMol team for their continued production of large line lists and pertinent data critical to high-temperature atmospheric modeling. EGN’s research was supported by an appointment to the NASA Postdoctoral Program at the NASA Ames Research Center, administered by Universities Space Research Association under contract with NASA. Finally, we thank the anonymous reviewers for their thoughtful comments.
\section{Software}
We used the publicly available $ExoCross$ code \citep{yurchenko2018exocross} developed by the ExoMol group to generate the cross sections, and the {\tt CHIMERA} radiative transfer and retrieval tool to produce the simulated emission/transmission spectra.

\clearpage

\appendix
\section{Appendices}
\subsection{Calculating the Lorentz coefficient ($\gamma_\mathrm{L}$) based on collision theory} \label{sec:A1}

The interaction between a given absorber with other atoms or molecules (i.e., perturbers) induces an extra width in the recorded spectral lines. In order to understand and calculate the line width, this interaction may be treated in two ways. The first approach is the classical approximation method in which the collision between the absorber and broadener occurs through a straight trajectory. The second method, however, uses quantum mechanical theory to describe the motion of perturbers. Limitations in both methods have been discussed in detail by \citet{Baranger1958}. In this study, we used Anderson theory which deals simultaneously with the motion of perturbers and the transition between energy levels with classical and quantum mechanics \citep{Anderson1949}. The total energy in the system is then solved through perturbation theory, and the collisional half-with of the spectral line can be represented by Eq.~\ref{eq:collision-HWHM} (see Eq.~28 by \citet{Anderson1949}):

\begin{equation}\label{eq:collision-HWHM}
    \Delta\nu_{\nicefrac{1}{2}} = \frac{1}{2 \pi} n_b \bar{v}_{th}\sigma_{col} 
\end{equation}

\noindent where $n_b$ is the number density of perturbers, $\bar{v}_{th}$ is the mean relative velocity of the gas (=$\sqrt{8k_BT / \mu \pi}$), and $\sigma_{col}$ is the collision cross-section in $\AA^2$ unit. Using the gas kinetic theory, HWHM linewidth $\Delta\nu_{1/2}$ (in Hz/Pa) may be written as:

\begin{equation}\label{eq:A2}
    \Delta\nu_{\nicefrac{1}{2}} =  \left(\frac{8k_B T}{\mu \pi}\right)^{1/2} \frac{1}{2 k_B \pi}  \ T^{-1/2} \ \sigma_{col} \ p_{b} 
\end{equation}
where $k_{\mathrm{B}}$ is a Boltzmann constant and $\mu$ is the reduced mass of the colliding pair in atomic mass unit ($\mu=M_{absorber}M_{broadener}/(M_{absorber}+M_{broadener}) $), and $p_{\mathrm{b}}$ is the partial pressure of the broadening gas. Comparing Eq.~\ref{eq:A2} to  Eq.~\ref{eq:Lorentz-HWHM} shows that $\Delta\nu_{\nicefrac{1}{2}}$ is equivalent to Lorentz HWHM ($\Gamma_\mathrm{L}$) (see Eq.~36c and its related discussion by \citet{Baranger1958}). By converting the unit of Eq.~\ref{eq:A2} to cm$^{-1}$/atm and cm$^{-1}$/bar , the Lorentz coefficient $\gamma_\mathrm{L}$  can be written as:

\begin{equation}\label{eq:A3}
    \gamma_\mathrm{L} [cm^{-1}/atm]=  0.0567 \  (T \mu)^{-1/2} \ \sigma_{col}
\end{equation}
or, 
\begin{equation}\label{eq:A4}
    \gamma_\mathrm{L} [cm^{-1}/bar]=  0.0574 \  (T \mu)^{-1/2} \ \sigma_{col}
\end{equation}
in which, the temperature dependence of the $\gamma_\mathrm{L} (T)$ is:  
\begin{equation}\label{eq:A5}
    \gamma_\mathrm{L}(T)= \gamma_\mathrm{L} (T_0) \left(\nicefrac{T}{T_0}\right)^{-1/2} 
\end{equation}

Note that this equation can only provide a rough estimate of the pressure-broadening coefficients because the accuracy of the results relies on the collision cross section and also kinetic theory to provide the temperature-dependence coefficient of -1/2.

\clearpage


\begin{table}[H]
\centering
  \footnotesize 
  \caption{Available calculated/estimated Lorentz coefficient, $\gamma_{\mathrm{L,b}}$[cm$^{-1}$/atm] from the literature (for Eq.~\ref{eq:Lorentz-HWHM}). }\label{tab:Lorentz-coefficient-known-data}
\centering 

\begin{tabular}{ l c l l l } 
\hline\hline
Absorber & Broadener   & $\gamma_{\rm{L,b}}(J=0)^{\color{red}{*}}$  & $\gamma_{\rm{L,b}}$ & Reference    \\
\hline        
\ce{H2O}  & \ce{H2}   &  0.09  & laboratory data & Ba17 \\
\ce{H2O}  & \ce{He}   &  0.02 & laboratory data &  Ba17\\
\hline
TiO, VO  & \ce{H2}   &  0.1      &  0.1 - 0.002$J_{\rm lower}$$^{\color{red}{**}}$ & Sh07 \\
TiO, VO  & \ce{He}   & 0.06  & 0.06 - 0.0012$J_{\rm lower}$  & \\
\hline        
FeH, TiH, CrH  & \ce{H2}   & 0.075    & 0.075 - 0.001$J_{\rm lower}$ & Du03,Bu02,Bu05 \\
FeH, TiH, CrH  & \ce{He}   &   0.045   &    0.045 - 0.0006$J_{\rm lower}$&               \\

\hline        
\hline
\end{tabular}
       \hfill\parbox[t]{\linewidth}{
        $^{\color{red}{*}}$$\gamma_{\mathrm{L,b}}$ is reported for $J$=0 and 296~K. For \ce{H2O}, we adopted the $J$-dependency proposed in the cited studies. The $J$-dependency of other molecules is discussed in \S\ref{sec:fitting-P-broadening}. \\ 
       $^{\color{red}{**}}$Note, this equation was mistakenly proposed to provide a FWHM of $\gamma_{\rm{L}}$ by \citet{Sharp2007}, but in fact, it is for calculating the HWHM value. \\ 
       References: Ba17: The \ce{H2O} Lorentz coefficients were collected by the ExoMol group
        \href{http://exomol.com/data/molecules/H2O/}{http://exomol.com/data/molecules/H2O/}, and was discussed by  \citet{Barton2017}; 
        Bu05: \citep{Burrows2005TiH};
        Bu03: \citep{Burrows2002_CrH};
        Du03: \citep{Dulick2003};
        Sh07: \citet{Sharp2007}. }
\end{table}    

\clearpage
\pagebreak

 \begin{table}
\centering
  \footnotesize 
  \caption{Collision cross-section used to calculate the Lorentz pressure-broadening coefficient, $\gamma_{\mathrm{L,b}}$[cm$^{-1}$/atm]  }\label{tab:collisionXS}   
\centering 
\begin{tabular}{ l c c l l l l } 

\hline\hline
Absorber & Broadener  &  $\mu$ & $\sigma_{col}$  & $\gamma_{\mathrm{L,b}}$ &   Reference    \\
\hline
$^{24}$Mg$^{1}$H  & \ce{H2}   &  1.85 &    NA      &0.08 & =$\gamma_\mathrm{L}$(CaH--\ce{H2}) \\
$^{24}$Mg$^{1}$H  & \ce{He}   &  3.45 & NA          &0.05 &=$\gamma_\mathrm{L}$(CaH--He)  \\
\hline        
$^{40}$Ca$^{1}$H  & \ce{H2}   &  1.91 &   NA        &0.08 & =1.6$\gamma_\mathrm{L}$(CaH--He)\\
$^{40}$Ca$^{1}$H  & \ce{He}   &  3.64 &  $\sim$27     &0.05 & [1] \\
\hline        
$^{27}$Al$^{1}$H  & \ce{H2}   &  1.87 &     NA      &0.08  &  =$\gamma_\mathrm{L}$(CO--\ce{H2})  \\
$^{27}$Al$^{1}$H  & \ce{He}   &  3.50 &     NA      &0.05 &   =$\gamma_\mathrm{L}$(CO--He) \\
\hline        
$^{28}$Si$^{16}$O  & \ce{H2}   &  1.91 & $\sim$20 & 0.05 & [2] \\
$^{28}$Si$^{16}$O  & \ce{He}   & 3.67 & $\sim$10 & 0.02 & [3] \\
\hline
\end{tabular}
       \hfill\parbox[t]{\linewidth}{
       Notes:\\
       1. The reported $\sigma_{\mathrm{col}}$ [$\AA^2$] is the excitation collision cross section (for $J=0\rightarrow1$ in most cases) for collision kinetic energy calculated for 296 K ($E_{kin}=(4k_BT/\pi)$ $\approx5.2\times 10^{-21}$Joule=262 \  cm$^{-1}$ = 0.032 eV).\\
       2. The calculated $\gamma_{\mathrm{L,b}}$ is used to calculate HWHM Lorentz width, $\Gamma_{\mathrm{L,b}}$ in Eq.~\ref{eq:Lorentz-HWHM}.\\ 
       3. $\gamma_{\mathrm{L,b}}$ is reported at temperature 296 K for $J$=0. \S\ref{sec:fitting-P-broadening} discusses its $J$-dependency.\\  
       4. In the case of AlH--\ce{H2} and AlH--He systems, we used CO Lorentz coefficients because they both have a very close dipole moment and reduced mass. \\
       5. In a case of MgH, CaH, and SiO molecules, we used this assumption of $\gamma_{\mathrm{L,b}}$(He)=0.6$\gamma_{\mathrm{L,b}}$(\ce{H2}).\\
       Reference: [1]\citep{Akpinar2010}, [2]\citep{Yang2018}, [3]\citep{Dayou2006}.}
\end{table}    

\clearpage
\pagebreak

\begin{table*} 
\centering
  \footnotesize 
  \caption{Fitted coefficients of the Fourth-order Pad\'e formula (see Eq.~\ref{eq:pade-fitting}) to provide $\gamma_{\mathrm{L}}$[cm$^{-1}$/atm] for $J$ quantum numbers up to 500.  }
  \label{tab:fitting-coefficients-Pade}   
\centering 
\begin{tabular}{ l c r r r r r r r r r r r  } 

\hline\hline
Absorber & Broadener   & a$_0$ & a$_1$  & a$_2$ &   a$_3$ & b$_1$  & b$_2$ &   b$_3$ &   b$_4$    \\
\hline
AlH & \ce{H2} & 7.6101e-02 & -4.3376e-02 & 1.9967e-02 & 2.4755e-03 & -5.6857e-01 & 2.7436e-01 & 3.6216e-02 & 1.5350e-05    \\
AlH & He & 4.8630e-02 & 2.1731e+03 & -2.5351e+02 & 3.8607e+01 & 4.4644e+04 & -4.4438e+03 & 6.9659e+02 & 4.7331e+00    \\
CaH, MgH & \ce{H2} & 8.4022e-02 & -8.2171e+03 & 4.6171e+02 & -7.9708e+00 & -9.7733e+04 & -1.4141e+03 & 2.0290e+02 & -1.2797e+01    \\
CaH, MgH & He & 5.0424e-02 & -1.1014e+02 & 2.2833e+01 & 2.8527e-01 & -2.4407e+03 & 5.5472e+02 & -3.4490e+01 & 6.5808e+00    \\
CrH, FeH, TiH & \ce{H2} & 7.0910e-02 & -6.5083e+04 & 2.5980e+03 & -3.3292e+01 & -9.0722e+05 & -4.3668e+03 & 6.1772e+02 & -2.4038e+01    \\
CrH, FeH, TiH & He & 4.2546e-02 & -3.0981e+04 & 1.2367e+03 & -1.5848e+01 & -7.1977e+05 & -3.4645e+03 & 4.9008e+02 & -1.9071e+01    \\
SiO & \ce{H2} & 4.7273e-02 & -2.7597e+04 & 1.1016e+03 & -1.4117e+01 & -5.7703e+05 & -2.7774e+03 & 3.9289e+02 & -1.5289e+01    \\
SiO & He & 2.8364e-02 & -6.7705e+03 & 2.7027e+02 & -3.4634e+00 & -2.3594e+05 & -1.1357e+03 & 1.6065e+02 & -6.2516e+00    \\
TiO, VO & \ce{H2} & 1.0000e-01 & -2.4549e+05 & 8.7760e+03 & -8.7104e+01 & -2.3874e+06 & 1.6350e+04 & 1.7569e+03 & -4.1520e+01    \\
TiO, VO & He & 4.0000e-02 & -2.8682e+04 & 1.0254e+03 & -1.0177e+01 & -6.9735e+05 & 4.7758e+03 & 5.1317e+02 & -1.2128e+01    \\

\hline
\end{tabular}
       \hfill\parbox[t]{\linewidth}{
       Notes:\\
       1. See \S\ref{sec:fitting-P-broadening} for a detailed discussion regarding the calculation of these Fourth-order Pad\'e coefficients. \\
       2. These coefficients were used to compute the pre-generated absorption cross-sections in this study.}

\end{table*}    

\clearpage
\pagebreak

\begin{table*}[!ht]
  \footnotesize 
  \caption{Summary list of opacities: molecules, temperature, pressures, and their line lists (see \S\ref{sec:linelist} $\&$ \ref{sec:Results}).}\label{tab:Summary-linelist}   
\centering 
\begin{tabular}{ l l l l c c c c c l } 
\hline\hline
Absorber	& Line list	 &	$\lambda$ [$\mu$m]  & No. Lines& P [bar] & T [K] & PT points & I$_{cutoff}$ & Abundance($\%$) & Reference\\
\hline
$^{26}$Al$^{1}$H  &  WYLLoT   &     0.4--330           & 3.6$\times 10^4$   & 10$^{-6}$--3000&75-4000&1460& 10$^{-50}$ & 0.00 &Yu18     \\ 
$^{27}$Al$^{1}$H  &  WYLLoT   &     0.4--330           & 3.6$\times 10^4$   & 10$^{-6}$--3000&75-4000&1460& 10$^{-50}$&  100.00 & Yu18     \\ 
$^{27}$Al$^{2}$H  &  WYLLoT   &     0.4--330           & 3.6$\times 10^4$   & 10$^{-6}$--3000&75-4000&1460& 10$^{-50}$&  0.00 & Yu18     \\ 
\hline
$^{40}$Ca$^{1}$H  &  MoLLIST	   &     0.45-1, 2-14        & 1.9$\times 10^4$   & 10$^{-6}$--3000&75-4000&1460& 10$^{-50}$&  96.93 & Al17,Li12 \\
$^{40}$Ca$^{1}$H  &  Yadin	   &     1-2, 14-330      &    2.7$\times 10^4$   & 10$^{-6}$--3000&75-4000&1460& 10$^{-50}$&  96.93 & Ya12  \\
\hline
$^{52}$Cr$^{1}$H  &  MoLLIST	   &     0.7--1.6      &    1.4$\times 10^4$   & 10$^{-6}$--3000&75-4000&1460& 10$^{-50}$&  83.78 & Bu02 \\
\hline
$^{56}$Fe$^{1}$H  &  MoLLIST	   &     0.37--330     &     1.2$\times 10^5$   & 10$^{-6}$--3000&75-4000&1460& 10$^{-50}$&  91.74 & We10\\
\hline

$^{24}$Mg$^{1}$H &  MoLLIST	   &     0.35--1.2        & 3.1$\times 10^4$  & 10$^{-6}$--3000&75-4000&1460& 10$^{-50}$& 78.99 &  Gh13 \\
$^{24}$Mg$^{1}$H &  Yadin	   &     1.2--330        & 6.7$\times 10^3$   & 10$^{-6}$--3000&75-4000&1460& 10$^{-50}$& 78.99 &  Ya12 \\
$^{25}$Mg$^{1}$H &  Yadin	   &     1.2--330        & 6.7$\times 10^3$   & 10$^{-6}$--3000&75-4000&1460& 10$^{-50}$& 10.00 &  Ya12 \\
$^{26}$Mg$^{1}$H &  Yadin	   &     1.2--330        & 6.7$\times 10^3$   & 10$^{-6}$--3000&75-4000&1460& 10$^{-50}$& 11.01 &   Ya12 \\

\hline
$^{28}$Si$^{16}$O  &  EBJT	   &     1.67--330                  &   1.8$\times 10^6$   & 10$^{-6}$--3000&75-4000&1460& 10$^{-50}$& 92.00 & Ba13 \\
$^{28}$Si$^{17}$O  &  EBJT	   &     1.67--330                  &   1.8$\times 10^6$   & 10$^{-6}$--3000&75-4000&1460& 10$^{-50}$& 0.03 & Ba13 \\
$^{28}$Si$^{18}$O  &  EBJT	   &     1.67--330                  &   1.8$\times 10^6$   & 10$^{-6}$--3000&75-4000&1460& 10$^{-50}$&  0.19 & Ba13 \\
$^{29}$Si$^{16}$O  &  EBJT	   &     1.67--330                  &   1.8$\times 10^6$   & 10$^{-6}$--3000&75-4000&1460& 10$^{-50}$&  4.67 & Ba13 \\
$^{30}$Si$^{16}$O  &  EBJT	   &     1.67--330                  &   1.8$\times 10^6$   & 10$^{-6}$--3000&75-4000&1460& 10$^{-50}$&  3.08 & Ba13 \\
\hline
$^{48}$Ti$^{1}$H  &  MoLLIST	   &     0.4--2        &  2.0$\times 10^5$   & 10$^{-6}$--3000&75-4000&1460& 10$^{-50}$&  73.71 & Bu05   \\
\hline
$^{46}$Ti$^{16}$O  &  TOTO	   &     0.3--330                  &   3.0$\times 10^7$   & 10$^{-6}$--3000&75-4000&1460& 10$^{-50}$& 8.22 & Mc19 \\
$^{47}$Ti$^{16}$O  &  TOTO	   &     0.3--330                  &   3.0$\times 10^7$   & 10$^{-6}$--3000&75-4000&1460& 10$^{-50}$& 7.42 & Mc19 \\
$^{48}$Ti$^{16}$O  &  TOTO	   &     0.3--330                  &   3.0$\times 10^7$   & 10$^{-6}$--3000&75-4000&1460& 10$^{-50}$& 73.54&  Mc19 \\
$^{49}$Ti$^{16}$O  &  TOTO	   &     0.3--330                  &   3.0$\times 10^7$   & 10$^{-6}$--3000&75-4000&1460& 10$^{-50}$& 5.40 & Mc19 \\
$^{50}$Ti$^{16}$O  &  TOTO	   &     0.3--330                  &   3.0$\times 10^7$   & 10$^{-6}$--3000&75-4000&1460& 10$^{-50}$& 5.17 & Mc19 \\
\hline

$^{51}$V$^{16}$O  &  VOMYT	   &     0.28--330                  &   2.8$\times 10^8$    & 10$^{-6}$--3000&75-4000&1460& 10$^{-50}$& 99.5 & Mc16 \\
\hline

$^{1}$H$_2^{16}$O  &  POKAZATEL	   &     0.25--100                  &   1.1$\times 10^9$    & 10$^{-6}$--300&500-4000&1460& 10$^{-34}$&  99.75 & Po18 \\
\hline
\hline
\end{tabular}
\centering 
       \scriptsize
       \hfill\parbox[t]{\linewidth}{
       Notes:\\
       1. We used a single line list for some molecules such as TiO, SiO, and \ce{H2O}. For some molecules such as CaH and MgH, multiple line lists were used in order to compute opacity data for the full spectral range. See \S\ref{sec:linelist} and \S\ref{sec:Results} for more details on the line lists.  \\
        2. Species are weighted according to their solar/natural elemental abundances provided by NIST database: \href{https://www.nist.gov/pml/atomic-weights-and-isotopic-compositions-relative-atomic-masses}{www.nist.gov/pml/atomic-weights-and-isotopic-compositions-relative-atomic-masses}. \\
        3. For each species, absorption cross section data are provided in two ways: separate for each isotopologues and weighted based on their natural abundance. \\
       References: 
       Al17=\citet{Alavi2017CaH},
       Bu02=\citet{Burrows2002_CrH}
       Ba13=\citet{Barton2013SiO}, 
       Bu05=\citet{Burrows2005TiH}, 
       Gh13=\citet{GharibNezhad2013MgH},
       Li12=\citet{Li2012CaH},
       Mc19=\citet{McKemmish2019TiO-TOTO}, 
       Mc16=\citep{McKemmish2016}, 
       Po18=\citet{Polyansky2018}, 
       We10=\citet{Wende2010},
       Ya12=\citet{Yadin2012MgH-CaH}, 
       Yu18=\citet{Yurchenko2018AlH}}  

\end{table*}    

\clearpage
\pagebreak

\begin{table}[!htb]
  \centering 
	\scriptsize   
  \caption{Opacity computational details}\label{tab:comp}     
     \begin{tabular}{llllllllll}
       \toprule[\heavyrulewidth]\toprule[\heavyrulewidth]
        \multicolumn{5}{l}{\bf Absorbers:}\\
        & \ce{H2O} & TiO& VO & SiO & CaH \\
        & MgH & AlH  & TiH  & CrH & FeH\\
	\midrule[0.3pt]
        \multicolumn{7}{l}{{\bf Broadeners$^{\color{red}{*}}$}: 85\% \ce{H2} + 15\% He}\\
	\midrule[0.3pt]
     \multicolumn{7}{l}{\bf T[K] grid: (73 points)$^{\color{red}{**}}$}\\
    &  75 & 100  & 110 & 120 & 130 &  140 & 150 & 160  \\
    & 170 &  180 &  190  &  200 & 210 &  220 & 230 & 240 \\     
    &  250&  260 & 270 &  275 & 280 & 290  & 300 &  310\\ 
     &  320 & 330&  340 & 350 &  375 & 400  & 425 &  450 \\ 
    &  475 & 500&  525 & 550 &  575 & 600  & 650 &  700 \\ 
    &  750 & 800&  850 & 900 & 950 &  1000 & 1100 & 1200\\ 
     & 1300 & 1400&  1500 &  1600 & 1700 & 1800 &  1900 &  2000 \\
      & 2100 & 2200 &  2300&  2400 & 2500 &  2600 &  2700 & 2800 \\
      & 2900 &  3000 &  3100& 3200 &  3300 &  3400 & 3500 & 3750 \\
    & 4000 \\
	\midrule[0.3pt]
     \multicolumn{7}{l}{\bf P[bar] grid: (20 points)}\\
 & 10$^{-6}$ & 3$\times$10$^{-6}$ & 10$^{-5}$ & 3$\times$10$^{-5}$ & 10$^{-4}$ & 3$\times$10$^{-4}$ \\
  & 10$^{-3}$ & 3$\times$10$^{-3}$ & 10$^{-1}$  & 3$\times$10$^{-1}$ & 10$^{-2}$ & 3$\times$10$^{-2}$    \\ 
 & 1 & 3  & 10 & 30  &  100 &   300  \\
 & 1000& 3000\\
 	\midrule[0.3pt]
   \multicolumn{6}{l}{ \bf Wing cutoff:$^{\color{red}{\dagger}}$}\\
      & \multicolumn{7}{l}{P$>$200 bar:  150 cm$^{-1}$} \\
      & \multicolumn{7}{l}{P$\leqslant$200 bar: 30 cm$^{-1}$} \\	
      \midrule[0.3pt]
      \multicolumn{7}{l}{\bf Opacity code: {\tt{ExoCross}}\citep{yurchenko2018exocross}}\\
      \midrule[0.3pt]
      \multicolumn{7}{l}{\bf Voigt Algorithm: \citet{Humlicek1979}}\\
    \bottomrule[\heavyrulewidth] 
   \end{tabular}
       \hfill\parbox[t]{8.3cm}{ 
       $^{\color{red}{*}}$These broadening mixture is a good choice for hydrogen-dominated atmospheres such as (Ultra) hot-Jupiters and M dwarfs. For high metallicity atmospheres such as  super-Earths and mini-Neptunes, better choice of broadeners such as \ce{H2O} and \ce{CO2} is essential.\\ 
       $^{\color{red}{**}}$ For \ce{H2O} molecule, absorption cross-sections are calculated for temperature range 500--4000K.\\
       $^{\color{red}{\dagger}}$To compute the line-profile and include the effect of pressure-broadening, Lorentz profile is used. The calculation of Lorentz profile for each individual line can be extended to the full spectral range or be limited to a few wavenumbers. This is one main challenge in computing opacities, and some studies such as \citep{Hartmann2002wing} and \citep{bezard2011} have shown the non-Lorentzian behavior of the line profile. See \S\ref{sec:opacity-challenge} for more discussion.}
\end{table}

\clearpage
\pagebreak

\begin{table*}[!htb]
  \footnotesize 
  \caption{Comparing different \ce{H2O} line lists.}\label{tab:H2O-linelists}   
   \centering 
\begin{tabular}{ l c c c c l c } 
\hline
Line Lists & Method   &  $\lambda$[ $\micron$ ]  &  $J_{max}$ &  No. Lines &Note& Ref. \\
\hline\hline
PS97 & $ab\  initio$ &	 0.4 -- 100  	&  55 & $\sim$3$\times$10$^{8}$ & more accurate than BT2 for $\lambda >$1 & [1]   \\
\hline
BT2 & $ab\ initio$  &  0.33 -- 100  &	50 & $\sim$5$\times$10$^{8}$ & more transition than Ames1997, shorter wavelength & [2]\\ 
\hline
POKAZATEL & $ab\  initio$ 		& 0.25 -- 100   &  72&  $\sim$1$\times$10$^{9}$	& higher   intensity than BT2 at $\lambda <$1~$\micron$  & [3] \\ 
\hline
HITEMP & lab/$ab\  initio$	 & 0.33 -- 100  & 50 & $\sim$1$\times$10$^{8}$	& included strong BT2 lines and experimental lines$^{\color{red}{*}}$  & [4]\\ 
\hline
\end{tabular}
       \scriptsize
       \hfill\parbox[t]{\linewidth}{
       $^{\color{red}{*}}$After rigorous assessment of the impact of  the BT2 intensity at 296--4000K, \citet{Rothman2010} have kept only 25$\%$ of the BT2 lines (see their Eq.~2 and their criteria for different wavenumber ranges and temperatures).   \\
       Ref.: [1] \citep{Partridge1997}, [2] \citep{Barber2006}, [3] \citep{Polyansky2018-Water-POKAZATEL}, [4] \citep{Rothman2010} }  

\end{table*}    

\clearpage
\pagebreak

\begin{table*}[!htb]
  \footnotesize 
\centering 
  \caption{Summary of the current opacity databases for the molecules in our investigation.}\label{tab:Summary-compareXS}   
\centering 
\begin{tabular}{ l l l l l l l l } 
\hline\hline
Abs. Cross Sec.	& Linelist	 &	$\lambda$ [$\mu$m]  & T[K]& P[bar] & Points/cm$^{-1}$ & P-T Pairs & Wing cut-off [cm$^{-1}$] \\
\hline
\multicolumn{3}{l}{\bf H2O}\\
EXOPLINES  &  POKAZATEL  &  0.25--100   &   500--4000  & 10$^{-6}$--300 & 50--120 & 486 & 100--300 \\
Gandhi2020$^{\color{red}{*}}$  &  POKAZATEL  &  0.95--5      &  400--1600 & 10$^{-5}$--10 & 100 & 49 & Not reported\\
Malik2019$^{\color{red}{**}}$  &  BT2  &  0.5--20      &   50--2900 &10$^{-6}$--10$^{3}$ &  100 & 812 & 100 \\
Goyal2018$^{\color{red}{\dagger}}$  &  BT2  &  0.33--100      &  70--3000 & 10$^{-9}$--10$^{3}$ & 1000 & 800 & 100\\
MacDonald2019$^{\color{red}{\ddagger}}$  &  POKAZATEL  &  0.4--50      &  100--3500 & 10$^{-6}$--10$^{2}$ & 100 & 162 & 30\\
Freedman2014$^{\color{red}{\dagger\dagger}}$ &  Schwenke& 0.5--100 & 75--4000 & 10$^{-6}$--300 & 80-120 & 1060 & 25--250\\
Chubb2020$^{\color{red}{\ddagger\ddagger}}$ &  POKAZATEL& 0.3--50 & 100--3400 & 10$^{-5}$--100 & 2-75 & 594 & 10-500\\
\hline
\multicolumn{3}{l}{\bf TiO}\\
EXOPLINES  &  TOTO	   &     0.3--100        &   75--4000 &10$^{-6}$--3000 & 2--285 & 1460 & 30,150 \\
Malik2019  & Plez/Schwenke &  0.5--5      &  50--2900 &10$^{-6}$--10$^{3}$ & 100 & 812 & 100 \\
Goyal2018  &  Plez  &  0.3--100      &  70--3000 & 10$^{-9}$--10$^{3}$ & 1000 & 800 & 100\\
MacDonald2019  &  TOTO  &  0.4--50      &  100--3500 & 10$^{-6}$--10$^{2}$ & 100 & 162 & 30\\
Freedman2014 &  Schwenke/Allard& 0.32--100 & 75--4000 &10$^{-6}$--300 & 80-120 & 1060& 25--250 \\
Chubb2020 &  TOTO& 0.3--50 & 100--3400 & 10$^{-5}$--100 & 2-75 & 594 & 10-500\\
\hline
\multicolumn{3}{l}{\bf VO}\\
EXOPLINES   &  VOMYT	   &     0.28--100        &  75--4000 &10$^{-6}$--3000 & 2--285 & 1460 & 30,150 \\
Malik2019  &  VOMYT  &  0.5--20      &  50--2900 &10$^{-6}$--10$^{3}$ & 100 & 812 &100 \\
Goyal2018  &  VOMYT  &  0.3--100      &  70-3000 & 10$^{-9}$--10$^{3}$ & 1000 & 800 & 100\\
MacDonald2019  &  VOMYT  &  0.4--50      &  100--3500 & 10$^{-6}$--10$^{2}$ & 100 & 162 & 30\\
Freedman2014 &  Plez& 0.8--100 & 75--4000 &10$^{-6}$--300 & 80-120 & 1060 & 25--250\\
Chubb2020 &  VOMYT& 0.3--50 & 100--3400 & 10$^{-5}$--100 & 2-75 & 594 & 10-500\\
\hline
\multicolumn{3}{l}{\bf FeH}\\
EXOPLINES &   Dulick	   &     0.37--100     &     75--4000 &10$^{-6}$--3000 & 2--285 & 1460 & 30,150 \\
Goyal2018  &  Wende2010  &  0.67--50      &  70--3000 & 10$^{-9}$--10$^{3}$ & 1000 & 800 & 10\\
MacDonald2019  &  Wende2010  &  0.4--50      &  100--3500 & 10$^{-6}$--10$^{2}$ & 100 & 162 & 30\\
Freedman2014 &  Dulick/Hargreaves& 0.37--100 & 75--4000 & 10$^{-6}$--300 & 80--120 & 1060 & 25--250\\
Chubb2020 &  Wende2010& 0.3--50 & 100--3400 & 10$^{-5}$--100 & 2-75 & 594 & 10-500\\
\hline
\multicolumn{3}{l}{\bf MgH}\\
EXOPLINES &  MoLLIST/Yadin	   &     0.35--100   & 75--4000 &10$^{-6}$--3000 & 2--285 & 1460 & 30,150 \\
Malik2019 &  Yadin	   &     1.2--20        & 50--2900 &10$^{-6}$--10$^{3}$ &  100 & 812 & 100 \\
MacDonald2019  &  MoLLIST/Yadin  &  0.4--50      &  100--3500 & 10$^{-6}$--10$^{2}$ & 100 & 162 & 30\\
Freedman2014 &  Weck& 0.3--2.5 & 75--4000 &10$^{-6}$--300 & 80--120 & 1060 & 25--250\\
Chubb2020 &  Wende2010& 0.3--50 & 100--3400 & 10$^{-5}$--100 & 2-75 & 594 & 10-500\\
\hline
\multicolumn{3}{l}{\bf AlH}\\
EXOPLINES  &  WYLLoT   &     0.4--100         & 75--4000 &10$^{-6}$--3000 & 2--285 & 1460 & 30,150 \\ 
Malik2019 &  WYLLoT	   &     0.4--20        & 50--2900 &10$^{-6}$--10$^{3}$ &   100 & 812 &100\\
MacDonald2019  &  WYLLoT  &  0.4--50      &  100--3500 & 10$^{-6}$--10$^{2}$ & 100 & 162 & 30\\
Chubb2020 &  WYLLoT& 0.3--50 & 100--3400 & 10$^{-5}$--100 & 2-75 & 594 & 10-500\\
\hline
\multicolumn{3}{l}{\bf CaH}\\
EXOPLINES  &  MoLLIST/Yadin	   &     0.45-100        & 75--4000 &10$^{-6}$--3000 & 2--285 & 1460 & 30,150 \\
Malik2019 &  Yadin	 & 1--20 & 50--2900  & 10$^{-6}$--10$^{3}$       &  100 & 812 &100 \\
MacDonald2019  &  Yadin  &  1--50      &  100--3500 & 10$^{-6}$--10$^{2}$ & 100 & 162 & 30\\
Freedman2014 &  Weck& 0.8--2.5 & 75--4000 & 10$^{-6}$--300 & 80--120 & 1060 & 25--250\\
Chubb2020 &  MoLLIST& 0.3--50 & 100--3400 & 10$^{-5}$--100 & 2-75 & 594 & 10-500\\

\hline
\end{tabular}
\centering 
       \scriptsize
       \hfill\parbox[t]{\linewidth}{
       $^{\color{red}{*}}$Gandhi2020=\citet{Gandhi2020XS-HighRes} ACSs data were  generate to be used for calculating cross-correlation function for high-resolution exoplanet spectroscopy.\\ 
       $^{\color{red}{**}}$Malik2019=\citet{Malik2019SelfLuminous} ACS data have been used to model the irradiated exoplanets. The pressure grid used for the opacities goes from $10^{-6}$ bar to $10^{3}$ bar with a logarithmic step size of 1/3 dex, that makes 28 pressure points. The temperature grid goes from 50 K to 2900 K with a linear step size of 50 K between 50 K and 700 K, a step size of 100 K between 700 K and 1500K, and a step size of 200 K between 1500 K and 2900 K. That is 29 temperature points in total. Opacities above 3000 K were extrapolate. \\
       $^{\color{red}{\dagger}}$Goyal2018=\citet{Goyal2018ATMO} ACS data have been used in {\tt ATMO} code for modeling forward exoplanet spectra.\\
       $^{\color{red}{\ddagger}}$MacDonald2019=\citet{macdonald2019POSIEDON} (see chapter 5) ACS data are a part of {\tt POSIEDON} database for atmospheric retrievals.\\
       $^{\color{red}{\dagger\dagger}}$Freedman2014=\citep{Lupu2014EarthLike, Freedman2008, Freedman2014} were used in  {\tt Exo-Transmit} and {\tt PICASO} code \citep{Kempton2017ExoTransmit, Batalha2019Picaso} to model transmission, emission, and reflected exoplanetary atmospheric spectra.\\
       $^{\color{red}{\ddagger\ddagger}}$Chubb2020=\citet{Chubb2020}: In this work, the resolving power is 15,000 and the wing cut-off is 500$\times \gamma_{\mathrm{Voigt}}$. In comparison, our resolution is 2-20 times larger in order to fully and accurately could be employed in modeling $JWST$- and high-resolution cross-correlation technique. Our wing cut-off is also a constant number of 30 and 150 cm$^{-1}$ because the real shape of Voigt profile is non-Lorentzian and using a very large number can lead to an intensive overestimation of the opacity continuum. }  
  
\end{table*}    

\clearpage
\pagebreak

\clearpage
\pagebreak

\clearpage

\pagebreak

\clearpage
\pagebreak

\pagebreak

\clearpage

\pagebreak

\clearpage

\pagebreak

\clearpage
\pagebreak

\clearpage

\pagebreak

\clearpage
\pagebreak

\clearpage

\pagebreak

\pagebreak

\pagebreak

\pagebreak

\pagebreak

\pagebreak

\clearpage
\pagebreak

\clearpage
\pagebreak

\bibliography{reference_8}
\end{document}